\documentclass[%
 reprint,
 amsmath,amssymb,
aps%,longbibliography
]{revtex4-1}
\usepackage{url}
\usepackage[colorlinks,linkcolor=blue]{hyperref}
\usepackage{graphicx}% Include figure files
\usepackage{dcolumn}% Align table columns on decimal point
\usepackage{bm}% bold math
\begin{document}

\title{Resilient quantum gates on periodically driven Rydberg atoms}
\author{Jin-Lei Wu$^{1}$}\author{Yan Wang$^{1}$}\author{Jin-Xuan Han$^{1}$}\author{Shi-Lei Su$^{2}$}\author{Yan Xia$^{3}$}\author{Yongyuan Jiang$^{1}$}\email[]{jiangyy@hit.edu.cn}\author{Jie Song$^{1,4,5}$}\email[]{jsong@hit.edu.cn}

\affiliation{$^{1}$School of Physics, Harbin Institute of Technology, Harbin 150001, China}
\affiliation{$^{2}$School of Physics and Microelectronics, Zhengzhou University, Zhengzhou 450001, China}
\affiliation{$^{3}$Department of Physics, Fuzhou University, Fuzhou 350002, China}
\affiliation{$^{4}$Key Laboratory of Micro-Nano Optoelectronic Information System, Ministry of Industry and Information Technology, Harbin 150001, China}
\affiliation{$^{5}$Key Laboratory of Micro-Optics and Photonic Technology of Heilongjiang Province, Harbin Institute of Technology, Harbin 150001, China}

\begin{abstract}
Fault-tolerant implementation of quantum gates is one of preconditions for realizing quantum computation. The platform of Rydberg atoms is one of the most promising candidates for achieving quantum computation. We propose to implement a controlled-$Z$ gate on Rydberg atoms where an amplitude-modulated field is employed to induce Rydberg antiblockade. Gate robustness against the fluctuations in the Rydberg-Rydberg interaction can be largely enhanced by adjusting  amplitude-modulated field. Furthermore, we introduce a Landau-Zener-St\"{u}ckelberg transition on the target atom so as to improve the gate resilience to the deviation in the gate time and the drift in the pulse amplitude. With feasible experimental parameters, one can achieve the gate with low fidelity errors caused by atomic decay, interatomic dipole-dipole force, and Doppler effects. Finally, we generalize the gate scheme into multiqubit cases, where resilient multiqubit phase gates can be obtained in one step with an unchanged gate time as the number of qubits increases.
\end{abstract}
\maketitle

\section{Introduction}
As one of the most promising candidates of implementing quantum computation and simulating many-body physics~\cite{PhysRevLett.85.2208,Saffman2010,PhysRevLett.121.123603,Omran570}, Rydberg atoms have been paid increasing attentions due to the long lifetime of internal states and the state-dependent interaction properties~\cite{Gallagher1994}. The powerful Rydberg-Rydberg interaction~(RRI) allows at most one of atoms in a small volume excited to a Rydberg state, i.e., a ``Rydberg blockade"~\cite{Gallagher1994,PhysRevLett.87.037901}. This feature can be used to form a Rydberg superatom and enable the quantum information processing in a mesoscopic scale by storing quantum information in collective states of atomic ensembles~\cite{PhysRevLett.87.037901}. Furthermore, the Rydberg antiblockade (RAB) is also an important regime where two or more atoms can be excited simultaneously, and it has applications for gaining the quantitative strength information of the RRI~\cite{Ates2007}, constructing quantum gates~\cite{PhysRevLett.85.2208,PhysRevLett.124.033603}, and creating steady entangled states~\cite{PhysRevLett.111.033607,PhysRevA.92.022328}.

Pulse schemes for fast and robust implementations of quantum gates are required for fault-tolerant quantum computation~\cite{PhysRevLett.123.100501}. For Rydberg-blockade gates, the square-wave-based schemes can be fast~\cite{PhysRevApplied.7.064017,PhysRevApplied.11.044035}, but they usually have multistep operations that are sensitive to control errors, which will cause multiple accumulation of decoherence and errors, especially for the compositions of multiqubit gates. Adiabatic one-~\cite{PhysRevLett.100.170504,PhysRevA.101.062309} and multistep~\cite{PhysRevA.96.042306,PhysRevA.97.032701} schemes may be robust but they are slow inevitably. In recent years, there also are interesting works done for quantum gates by using RAB regime, especially for one-step quantum gates~\cite{PhysRevA.93.012306,PhysRevA.96.012328,PhysRevA.95.022319,WU2020126039,PhysRevA.96.042335,Wu2020OL,PhysRevA.101.012347}.
Under the RAB regime, however the quantum gates are usually sensitive to the fluctuations in the RRI strength. On the one hand, the doubly-excited state $|rr\rangle$ attends in evolution~\cite{PhysRevLett.85.2208,Zhu:19}, and the gradient of the RRI potential will cause a strong interatomic dipole-dipole force~(DDF) to induce mechanical motion and decoherence of the atoms~\cite{PhysRevLett.110.213005}. On the other hand, the emergence of the RAB usually demands the strict conditions among the RRI strength, the atomic transition detuning, and Rabi frequency~\cite{PhysRevA.93.012306,PhysRevA.96.012328,PhysRevA.95.022319,WU2020126039}. In addition, the RAB regime is usually based on the second- or higher-order dynamics which is typically slow~\cite{PhysRevA.96.042335,Wu2020OL,PhysRevA.101.012347}.

For coherent control of quantum dynamics, a wealth of quantum phenomena can be found in periodically driven Rydberg atoms~\cite{PhysRevLett.120.123204,PhysRevA.98.052324,li2019periodicallydriven,niranjan2020landauzener,Mallavarapu2020}. In this work, with a periodical amplitude-modulated field on a control atom, we propose to implement a controlled-$Z$~(CZ) gate. The CZ gate is based on the RAB regime that is realized through offsetting the RRI with the amplitude modulation of the field on the control atom. By adjusting the maximum amplitude of the amplitude-modulated field, the doubly-excited states $|rr\rangle$ can be suppressed so as to enhance the gate robustness against fluctuations in the RRI strength and decoherence caused by interatomic DDF. Furthermore, when a frequency modulation is introduced for the field on the target atom to induce a Landau-Zener-St\"{u}ckelberg~(LZS) transition~\cite{PhysRevA.80.063407,SHEVCHENKO20101}, the gate scheme will be of resilience to the deviation in the gate time and the drift in the pulse amplitude. Even if atomic decay, interatomic DDF, and Doppler effect are taken into account, low gate errors can be accessible using feasible experimental parameters. In addition, we show the generalization of the gate scheme. Resilient high-fidelity multiqubit phase gates can be attained with the gate time being independent of the number of qubits.

\begin{figure*}\centering
	\includegraphics[width=0.9\linewidth]{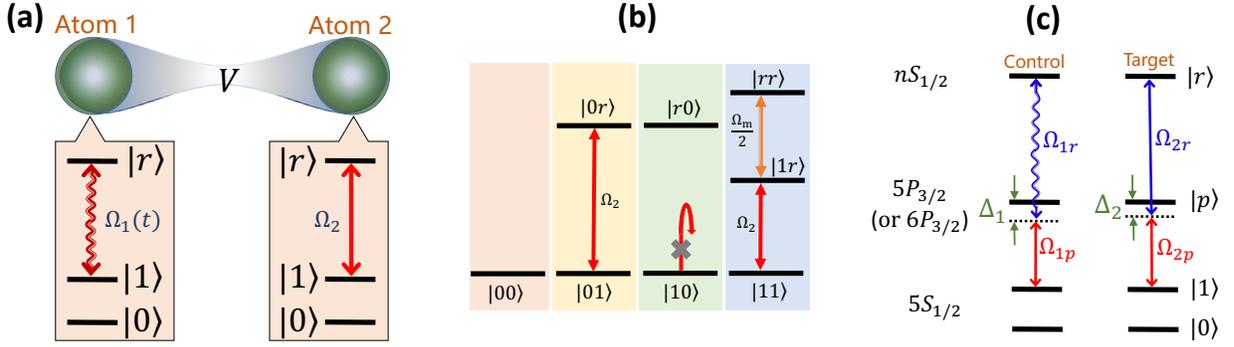}
	\caption{(a)~Sketch of two Rydberg atoms interacting with each other, atomic level structures and field-driven transitions. For Atom~1 called the control atom, the transition between the ground state $|1\rangle$ and the Rydberg state $|r\rangle$ is driven resonantly by a periodical amplitude-modulated field with a Rabi frequency $\Omega_1=\Omega_{\rm m}\cos(\omega t)$. For Atom~2, the target atom, the transition $|1\rangle\leftrightarrow|r\rangle$ is driven by a constant-amplitude field with a Rabi frequency $\Omega_2$. (b)~Transition dynamics of each initial state. $|00\rangle$ and $|10\rangle$ remain unchanged. $|01\rangle$ and $|11\rangle$ obey a Rabi transition and a Raman transition, respectively. (c)~Schematic diagram for the two-photon Rydberg excitation in two $^{87}$Rb atoms.}\label{f1}
\end{figure*}
The peculiarities of the present work can be concluded as following.
First, the previous RAB regimes are realized usually by using the second-~\cite{PhysRevLett.111.033607,PhysRevA.92.022328,PhysRevA.93.012306,PhysRevA.96.012328,PhysRevA.95.022319,WU2020126039,PhysRevA.96.042335,PhysRevA.101.012347} or higher-order~\cite{Wu2020OL,PhysRevA.98.032306,PhysRevA.101.012306} perturbation theory. The present RAB regime has the first-order dynamics, so it has great potential of applications for highly efficient Rydberg pumping and fast multiqubit gates.
Second, the RAB gates are suffering from common issues, extreme sensitivity to the fluctuations in the RRI strength~\cite{PhysRevA.93.012306,PhysRevA.95.022319,WU2020126039,PhysRevA.96.042335,Zhu:19,Wu2020OL,PhysRevA.98.032306} and fragile robustness against decoherence caused by interatomic DDF~\cite{PhysRevLett.110.213005}, which cause a great challenge for experimental implementations of high-fidelity quantum gates. In this work, these common issues are avoided. Besides, the LZS transition in the target atom makes the CZ gate more robust against error in the pulse area than other schemes based on fixed-area pulses~\cite{PhysRevApplied.7.064017,PhysRevApplied.11.044035}. Third, compared with the standard three-step Rydberg-blockade-based gate~\cite{PhysRevLett.85.2208} and its various variants~\cite{PhysRevApplied.9.051001,PhysRevA.96.042306,PhysRevA.97.032701,PhysRevA.98.052324,Shen:19,Liao:19,Liu2020} where an atom is excited before and de-excited after the excitation of the other atom, the present CZ gate can be implemented through just a Rabi cycle of only one two-atom computational product state, so it holds a transient shelving duration for Rydberg excitation, and suffers from less errors caused by atomic decay and Doppler dephasing due to atomic thermal motion. In addition, different from the one-step adiabatic Rydberg-blockade-based gate~\cite{PhysRevLett.85.2208} and its variants~\cite{PhysRevA.89.032334,PhysRevA.101.030301} that are relatively slow because of the limitation of the adiabatic criterion, the present CZ gate can be fast and thus experiences less accumulation of decoherence.
Finally, multiqubit gates can be achieved in one step, and the gate time does not increase with the number of qubits, different from recent Rydberg-blockade schemes of multiqubit gates that are achieved in multiple steps~\cite{PhysRevLett.123.170503,yu2020scalability} and RAB schemes where the gate time increases exponentially as the number of qubits increases~\cite{PhysRevA.98.032306,PhysRevA.101.012306}.

\begin{figure*}
	\includegraphics[width=0.66\linewidth]{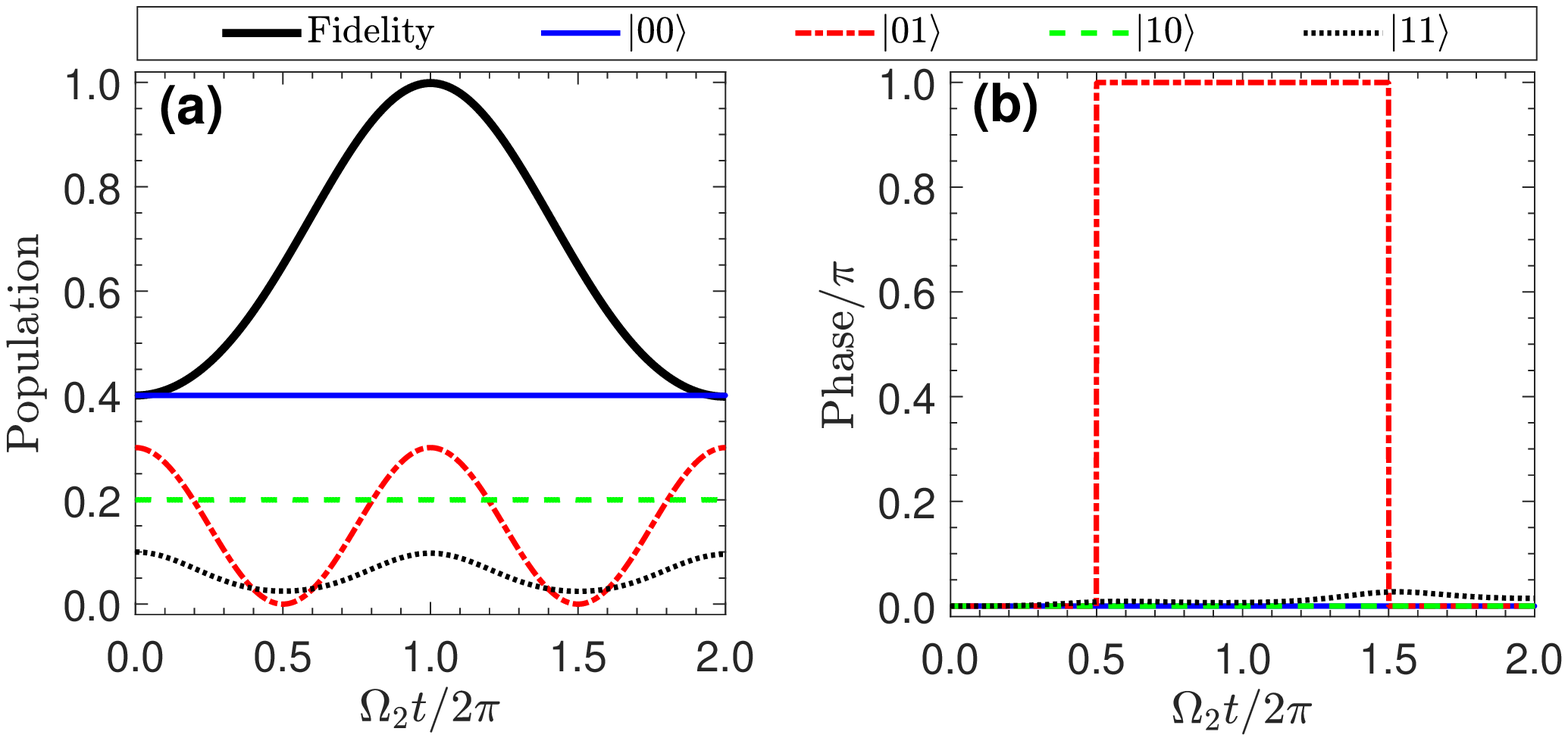}
	\caption{(a)~Population evolutions of the four computational states and the gate fidelity evolution. (b)~Phase evolutions of the four computational states. $\Omega_{\rm m}=2\sqrt3\Omega_2$ and $V=\omega=500\Omega_2$.}\label{f2}
\end{figure*}
\section{Rydberg antiblockade of two atoms}
As shown in Fig.~\ref{f1}(a), we consider that two neutral atoms are confined in two separated microscopic dipole traps, and interact with each other through the van der Waals interaction with a strength $V = C_6/d^6$ or the dipole-dipole interaction with $V = C_3/d^3$, depending on the interatomic distance $d$ and Rydberg states~\cite{Saffman2010,Gallagher1994}. $C_6$ and $C_3$ are the van der Waals and dipole-dipole interaction coefficients, respectively. In the interaction picture with the rotating-wave approximation, the Hamiltonian of the two-atom system is written as~($\hbar= 1$)
\begin{equation}\label{e1}
\hat{H}_{I}=\Big(\sum_{j=1,2}\frac{\Omega_j}2|1\rangle_j\langle r|+{\rm H.c.}\Big)+V|rr\rangle\langle rr|.
\end{equation}
The amplitude-modulated Rabi frequency $\Omega_1$ on Atom~1 (the control atom) is chosen as $\Omega_1=\Omega_{\rm m}\cos(\omega t)$, where $\Omega_{\rm m}$ is the maximum amplitude and $\omega$ the modulation
frequency. Such an amplitude-modulated field can be operated by an acousto-optic modulator with the help of an arbitrary waveform generator~\cite{dugan1997high}.

In order to illustrate the RAB effect, we consider first the two atoms prepared both initially in $|1\rangle$, then the two-atom Hamiltonian is
\begin{eqnarray}\label{e2}
\hat{H}_{11}&=&\frac{1}2[\Omega_{\rm m}\cos(\omega t)(|11\rangle\langle r1|+|1r\rangle\langle rr|)+\Omega_2(|11\rangle\langle 1r|\nonumber\\
&&+|r1\rangle\langle rr|)+{\rm H.c.}]+V|rr\rangle\langle rr|.
\end{eqnarray}
We define a rotating frame with a unitary operator $\exp(i\hat h_0t)$ and $\hat h_0\equiv\omega|rr\rangle\langle rr|$, so the Hamiltonian~(\ref{e2}) becomes
\begin{eqnarray}\label{e3}
\hat {\mathcal{H}}_{11}&=&\Big[\frac{\Omega_{\rm m}}4(e^{i\omega t}+e^{-i\omega t})|11\rangle\langle r1| +\frac{\Omega_{\rm m}}4(1+e^{-2i\omega t})\nonumber\\
&&\times|1r\rangle\langle rr|+\frac{\Omega_2}2(|11\rangle\langle 1r|+e^{-i\omega t}|r1\rangle\langle rr|)+{\rm H.c.}\Big]\nonumber\\
&&+(V-\omega)|rr\rangle\langle rr|.
\end{eqnarray}
Through neglecting highly frequent oscillations and higher-order couplings under the condition $\omega\gg\Omega_{\rm m},\Omega_2$, there exists the first-order dynamics with a three-state transition process $|11\rangle\leftrightarrow|1r\rangle\leftrightarrow|rr\rangle$ described by
an effective Hamiltonian
\begin{equation}\label{e4}
\hat {H}_{\rm eff}=\frac{\Omega_2}2|11\rangle\langle 1r|+\frac{\Omega_{\rm m}}4|1r\rangle\langle rr|+{\rm H.c.},
\end{equation}
for which the RAB condition $V=\omega$ has been used.

The Hamiltonian~(\ref{e4}) can enable the two-atom evolution from the initial state $|11\rangle$ to the doubly-excited state $|rr\rangle$ through a stimulated Raman process with the condition $|\Omega_{\rm m}|=2|\Omega_2|$. It exhibits an unconventional RAB regime, where the amplitude modulation of the field on the control atom plays a crucial role for offsetting the Rydberg blockade effect caused by the powerful RRI. This RAB regime can provide the first-order dynamics, different from the conventional ones that are based on the second- or higher-order dynamics~\cite{PhysRevA.93.012306,PhysRevA.96.012328,PhysRevA.95.022319,WU2020126039,PhysRevA.96.042335,Wu2020OL,PhysRevA.101.012347,PhysRevA.98.032306,PhysRevA.101.012306}. In the following, based on this RAB regime we show fast and resilient implementations of quantum gates, including two- and multiqubit phase gates.

\begin{figure*}
	\includegraphics[width=0.66\linewidth]{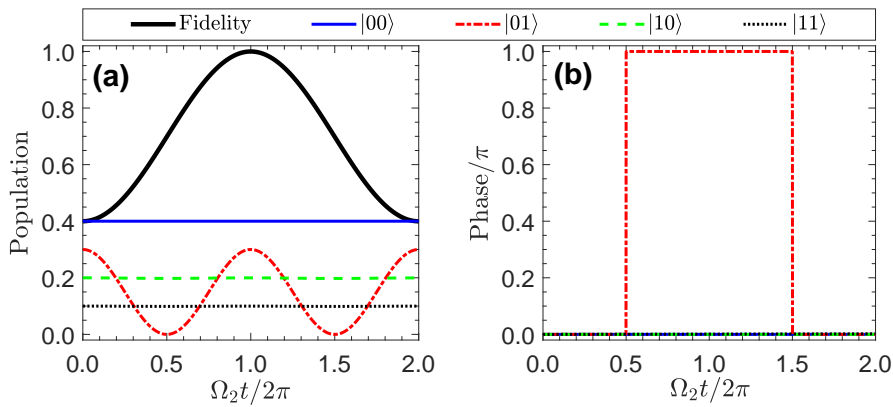}
	\caption{(a)~Population evolutions of the four computational states and the gate fidelity evolution; (b)~Phase evolutions of the four computational states. $\Omega_{\rm m}=100\Omega_2$ and $V=\omega=500\Omega_2$.}\label{f3}
\end{figure*}
\section{CZ gates on two Rydberg atoms}
\subsection{CZ gate combining cyclic Rabi and Raman transitions}
When the two-atom initial state is among computational states $|00\rangle$, $|01\rangle$ and $|10\rangle$, the RRI will not work. Because $|0\rangle$ is decoupled to the drive field there is no evolution for the initial state $|00\rangle$. For the initial state $|01\rangle$, the two-atom evolution is governed by the Hamiltonian
\begin{equation}\label{e5}
\hat {H}_{\rm 01}=\frac{\Omega_2}2|01\rangle\langle 0r|+{\rm H.c.},
\end{equation}
which describes a conditional Rabi oscillation of the Atom~2 with the condition that the Atom~1 is in $|0\rangle$. When the two atoms are in the initial state $|10\rangle$, the government Hamiltonian is
\begin{equation}\label{e6}
\hat {H}_{\rm 10}=\frac{\Omega_{\rm m}}2\cos(\omega t)|10\rangle\langle r0|+{\rm H.c.},
\end{equation}
which results in an evolutionary state $|\psi\rangle=\cos\theta(t)|10\rangle-i\sin\theta(t)|r0\rangle$ with $\theta(t)\equiv\Omega_{\rm m}\sin(\omega t)/{2\omega}$, according to Schr\"{o}dinger equation $i\partial |\psi\rangle/\partial t=\hat {H}_{\rm 10}|\psi\rangle$. Under the condition $\omega\gg\Omega_{\rm m}$, yielding $\theta(t)\simeq0$, the two atoms are frozen in $|10\rangle$. The transition dynamics of each initial state is visualized in Fig.~\ref{f1}(b).

Given that the computational states $|00\rangle$ and $|10\rangle$ remain unchanged during the evolution, to realize a CZ gate $\hat U_{\rm CZ}=|00\rangle\langle00|-|01\rangle\langle01|+|10\rangle\langle10|+|11\rangle\langle11|$, transformations $|11\rangle\mapsto|11\rangle$ and $|01\rangle\mapsto-|01\rangle$ should be accomplished synchronously. Dominated by the Hamiltonians~(\ref{e4}) and (\ref{e5}), the time-dependent probability amplitudes of initial states $|11\rangle$ and $|01\rangle$ can be calculated out, respectively, as
\begin{equation}\label{c11}
C_{11}(t)=\frac{\Omega_{\rm m}^2+4\Omega_2^2\cos\left(\frac{t}4\sqrt{\Omega_{\rm m}^2+4\Omega_2^2}\right)}{\Omega_{\rm m}^2+4\Omega_2^2},
\end{equation}
and
\begin{eqnarray*}
C_{01}(t)=\cos(t\Omega_2/2).
\end{eqnarray*}
Therefore, the synchronous transformations $|11\rangle\mapsto|11\rangle$ and $|01\rangle\mapsto-|01\rangle$ can be realized by performing a Rabi cycle concerning $|01\rangle$ and a cyclic Raman process concerning $|11\rangle$, respectively. The gate time is chosen as $T=2\pi/\Omega_2$, and the parameters are controlled with $\Omega_{\rm m}=2\sqrt3\Omega_2$.

For illustrating the performance of the CZ gate, based on the full Hamiltonian~(\ref{e1}) we simulate the population evolutions of the four computational states and the gate fidelity in Fig.~\ref{f2}(a) by choosing a superposition $|\psi_0\rangle=\sqrt{0.4}|00\rangle+\sqrt{0.3}|01\rangle+\sqrt{0.2}|1 0\rangle+\sqrt{0.1}|11\rangle$ as the initial state here and later for all discussions about the two-qubit CZ gates, where the gate fidelity is defined as $F=|\langle\Psi(t)|\hat U_{\rm CZ}|\psi_0\rangle|$ with $|\Psi(t)\rangle$ being the solution of the Schr\"{o}dinger equation $i\partial|\Psi(t)\rangle/\partial t=\hat H_I|\Psi(t)\rangle$. Also, we plot the phase evolutions of the four computational states in Fig.~\ref{f2}(b). Figure~\ref{f2} indicates that $|00\rangle$ and $|10\rangle$ are not changed, and with $T=2\pi/\Omega_2$, $|01\rangle$ and $|11\rangle$ evolve back to themselves but $|01\rangle$ gets a $\pi$ phase. Correspondingly, the gate fidelity reaches unity at the time $t=2\pi/\Omega_2$.

\subsection{CZ gate using a strong amplitude-modulated field on the control atom}
The gate of combining cyclic Rabi and Raman transitions is not robust. On the one hand, the RAB condition $V=\omega$ should be satisfied strictly. On the other hand, it is possible to exist a strong interatomic DDF that can induce mechanical motion and decoherence because the doubly-excited state $|rr\rangle$ may be occupied during the gate process. Therefore, we explore a more robust scenario where the two atoms are not excited simultaneously and the RAB condition can be loosened to some extent. To this end, it is a solution to suppress the evolution of $|11\rangle$, i.e., to remain $C_{11}(t)\simeq1$. From Eq.~(\ref{c11}) we know that $\Omega_{\rm m}\gg2\Omega_2$ is an advisable choice, which prevents the transition from $|11\rangle$.

The condition $V=\omega$ works only for the process $|1r\rangle\leftrightarrow|rr\rangle$, so under the condition $\Omega_{\rm m}\gg2\Omega_2$, it can be moderately relaxed and the gate is improved to be insensitive to the fluctuations in the RRI strength, which will be discussed later. We simulate the population evolutions of the four computational states and the gate fidelity in Fig.~\ref{f3}(a), and the phase evolutions of the four computational states in Fig.~\ref{f3}(b). Compared with Fig.~\ref{f2}(a), the population of $|11\rangle$ in Fig.~\ref{f3}(a) becomes unchanged. The phase evolutions in Fig.~\ref{f3}(b) of the four computational states are almost the same as Fig.~\ref{f2}(b), and the gate fidelity also reaches unity at $t=2\pi/\Omega_2$.\\

\begin{figure*}
	\includegraphics[width=0.66\linewidth]{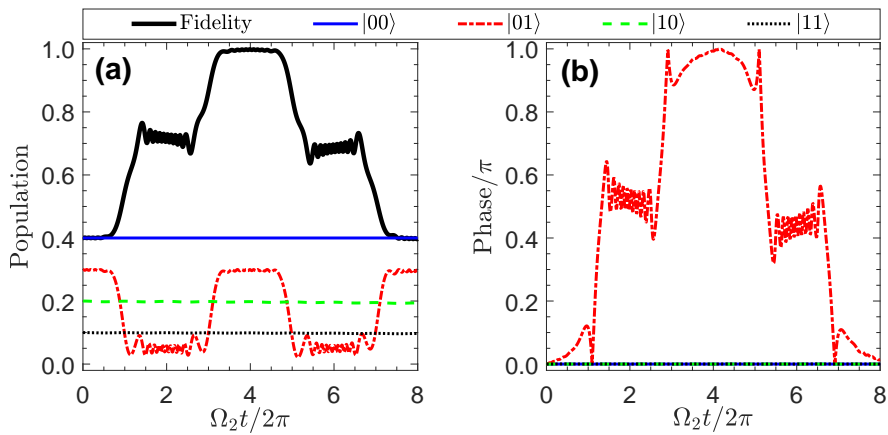}
	\caption{With the LZS transition on Atom~2, (a)~population evolutions of the four computational states and the gate fidelity evolution. (b)~Phase evolutions of the four computational states. $\bar\Delta=6\Omega_2$, $\Delta_0=5\Omega_2$, $\bar\omega=0.5\Omega_2$, $\Omega_{\rm m}=100\Omega_2$ and $V=\omega=500\Omega_2$.}\label{f4}
\end{figure*}
\subsection{CZ gate with an LZS transition}
In this subsection, we investigate an alternative robust CZ gate through introducing a frequency-modulated field to induce an LZS transition in the target atom. We change the constant-frequency field on Atom~2 into a frequency-modulated field so that the Rabi transition $|1\rangle\leftrightarrow|r\rangle$ is of a periodical detuning $\Delta=\Delta_0+\bar{\Delta}\cos(\bar{\omega}t)$ and becomes an LZS transition. Thus the full Hamiltonian of the two-atom system becomes
\begin{equation}\label{e8}
\hat{\mathcal{H}}_{I}=\Big(\sum_{j=1,2}\frac{\Omega_j}2|1\rangle_j\langle r|+{\rm H.c.}\Big)+\Delta|r\rangle_2\langle r|+V|rr\rangle\langle rr|.
\end{equation}
In the context of the condition $\Omega_{\rm m}\gg\Omega_2, (\Delta_0+\bar{\Delta})$, the evolution of the two-atom system can be governed by the following effective Hamiltonian
\begin{equation}\label{e9}
\hat {H}_{\rm e}=\hat {H}_{\rm 01}+\hat H_{\rm d}(t),
\end{equation}
where $\hat {H}_{\rm 01}$ is given by Eq.~(\ref{e5}) and $\hat H_{\rm d}(t)=[\Delta_0+\bar{\Delta}\cos(\bar{\omega}t)]|0r\rangle\langle0r|$. It will be convenient to move the Hamiltonian~(\ref{e9}) into the rotating frame with a unitary operator $\exp[i\int_0^t\hat H_{\rm d}(t')dt']$~\cite{PhysRevA.75.063414}, then $\hat {H}_{\rm e}$ becomes
\begin{equation}\label{e10}
\hat {H}'_{\rm e}=\frac{\Omega_2}2\sum_{n=-\infty}^\infty J_n(\frac{\bar\Delta}{\bar\omega})\exp[i(\Delta_0+n\bar\omega)t]|0r\rangle\langle 01|+{\rm H.c.},
\end{equation}
for which we have used the Jacobi-Anger expansion
$\exp\left[i\frac{\bar\Delta}{\bar\omega}\sin(\bar\omega t)\right]=\sum_{n=-\infty}^\infty J_n(\frac{\bar\Delta}{\bar\omega})\exp(in\bar\omega t)$
with $J_n(\cdot)$ denoting the $n$th-order Bessel function of the first kind.

The Hamiltonian~(\ref{e10}) can be comprehended as the two-level dynamics driven by polychromatic fields, each field corresponding to a Rabi frequency ${\Omega_2}J_n(\frac{\bar\Delta}{\bar\omega})$ and detuning $(\Delta_0+n\bar\omega)$, where the condition of the resonance field is $\Delta_0+n\bar\omega=0$. The dynamics strongly relies on the frequency-modulation parameters which determine whether or not a robust CZ gate can be implemented. For achieving a CZ gate, a full Rabi-like cycle between $|01\rangle$ and $|0r\rangle$ is supposed to be finished so as to make $|01\rangle$ evolve back to itself and get an extra $\pi$ phase. Therefore, a resonance field with the condition $\Delta_0+n\bar\omega=0$ is necessary to dominate the evolution. Besides, in order to make the gate robust against the deviation in the gate time and the drift in the pulse amplitude, there must be some near-resonance fields to interfere with the resonance field so that the gate can not be heavily damaged by the error in the resonance field, which requires $\bar\omega\sim\Omega_2/2$. Also, the resonance-field Rabi frequency should be as large as possible to ensure a fast operation.

Taking a set of frequency-modulation parameters $\{\bar\Delta=6\Omega_2,~\Delta_0=5\Omega_2,~\bar\omega=0.5\Omega_2\}$ as an example, the resonance-field Rabi frequency is $\Omega_{\rm r}=\Omega_2J_{-10}(12)=0.3{\Omega_2}$, so the operation time of implementing a CZ gate should be around $2\pi/\Omega_{\rm r}=3.33\times2\pi/\Omega_2$. However, because there exist about thirty non-negligible fields~($n\in[-25,5]$) that can work significantly or slightly for the CZ gate, the gate time may drift a little bit from $3.33\times2\pi/\Omega_2$. For more concrete illustration, based on the full Hamiltonian~(\ref{e8}) we simulate the population evolutions of the four computational states and the gate fidelity in Fig.~\ref{f4}(a), and the phase evolutions of the four computational states in Fig.~\ref{f4}(b). There is a wide time range $\Omega_2t/2\pi\in[3.5,4.5]$ when $|01\rangle$ evolves back to itself. Correspondingly, although during $\Omega_2t/2\pi\in[3.5,4.5]$ the phase of $|01\rangle$ is not stabilized thoroughly at the level of $\pi$, it deviates insignificantly by below $0.1\pi$. Therefore, a broad near-unity plateau of the gate fidelity appears during $\Omega_2t/2\pi\in[3.5,4.5]$, which will make the gate resistant to the deviation in the gate time $T=8\pi/\Omega_2$. In addition, because of the interference effect of polychromatic fields, the drift in the pulse amplitude of the frequency-modulated field will hardly affect the gate fidelity, which will be discussed in the next subsection.

\begin{figure*}
	\includegraphics[width=0.72\linewidth]{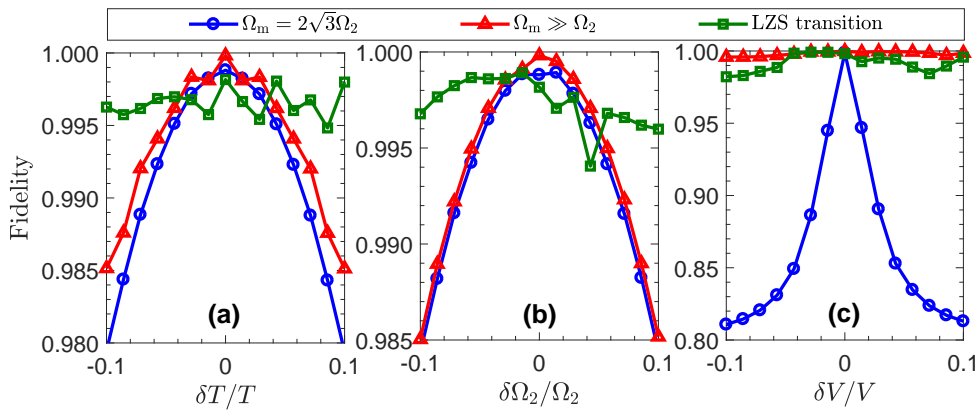}
	\caption{Influences on the gate fidelity of relative errors in (a)~gate time $T$, (b)~Rabi frequency $\Omega_2$, and (c)~RRI strength $V$. The gate time is $T=2\pi/\Omega_2$ for the case $\Omega_{\rm m}=2\sqrt3\Omega_2$ in Fig.~\ref{f2}, $T=2\pi/\Omega_2$ for $\Omega_{\rm m}\gg\Omega_2$ in Fig.~\ref{f3}, or $T=8\pi/\Omega_2$ for the LZS transition in Fig.~\ref{f4}.}\label{f5}
\end{figure*}
\subsection{Resilience to parameter inaccuracy}
In former subsections, it is foretold that the gate schemes with the condition $\Omega_{\rm m}\gg2\Omega_2$ have robustness to the fluctuations in the RRI strength. In particular, the scheme using the LZS transition is robust against the deviation in the gate time and the drift in the pulse amplitude. For further illustration, we investigate the influence on the gate fidelity of the inaccuracy of parameters including the gate time $T$, the amplitude of $\Omega_2$, and the RRI strength $V$.

We define a relative error $\delta X/X$ for a parameter $X$ with $\delta X$ being the deviation value from $X$. Then we simulate the influences on the gate fidelity of relative errors in $T$, $\Omega_2$, and $V$ in Figs.~\ref{f5}(a), (b), and (c), respectively. As foretold above, within the relative error range $\delta X/X\in[-0.1,0.1]$, the gate schemes with the condition $\Omega_{\rm m}\gg2\Omega_2$ are robust to the error in the RRI strength. In particular, errors in all the three parameters can not affect significantly the gate fidelity in the scheme of the LZS transition. On the contrary, the gate scheme of combining the cyclic Rabi and Raman transitions~($\Omega_{\rm m}=2\sqrt{3}\Omega_2$) is sensitive to errors in all the three parameters. For the error in $V$, it has a similar situation to the conventional RAB gate schemes where the gate fidelity has extreme sensitivity to the RRI strength fluctuations~\cite{PhysRevA.93.012306,PhysRevA.96.012328,PhysRevA.95.022319,WU2020126039,PhysRevA.96.042335,Wu2020OL,PhysRevA.101.012347,PhysRevA.98.032306,PhysRevA.101.012306}. In most schemes of Rydberg gates, the spatial positions of the interacting atoms are taken to be fixed, thus providing the fixed RRI strength. However, the atoms have finite temperatures and thus their positions can not be fixed actually due to residual thermal motion of the atoms~\cite{PhysRevLett.123.230501,PhysRevApplied.13.024059}, so it is scarcely possible to control accurately the RRI strength. Therefore, the present schemes with $\Omega_{\rm m}\gg2\Omega_2$ could be helpful for promoting robust quantum computation in neutral atoms.

\begin{figure}[b]
	\includegraphics[width=0.9\linewidth]{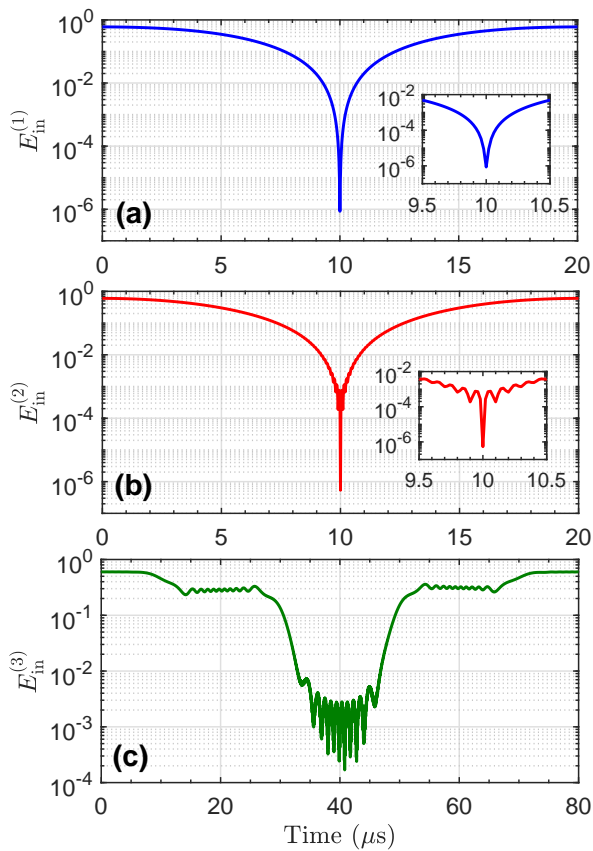}
	\caption{Intrinsic gate errors with the evolutionary gate time of the schemes based on (a)~$\Omega_{\rm m}=2\sqrt{3}\Omega_2$, (b)~$\Omega_{\rm m}/2\pi=10$~MHz without the LZS transition, and (c)~$\Omega_{\rm m}/2\pi=10$~MHz with the LZS transition. $V/2\pi\sim70$~MHz and $\Omega_2/2\pi=0.1$~MHz.}\label{f6}
\end{figure}
\section{Experimental considerations}
For the implementation of the phase gates in a concrete experiment, the Rydberg excitation from the ground state $|1\rangle$ to the Rydberg
state $|r\rangle$ can be achieved by a two-photon process~\cite{PhysRevLett.121.123603,PhysRevLett.123.170503,PhysRevLett.123.170503} in Rb atoms or a single-photon process~\cite{jau2016entangling} in Cs atoms. In the schemes of suppressing the evolution from $|11\rangle$, because of the condition $V\gg\Omega_{\rm m}\gg2\Omega_2$ a strong RRI strength that is chosen as over three orders of magnitudes larger than $\Omega_2$ for numerical simulations. Therefore, a Rydberg state with a sufficiently large principal quantum number or a relatively small interatomic separation is desired so as to ensure a strong RRI strength~\cite{Saffman2010}. Here we consider $^{87}$Rb atoms interacting with each other through the van der Waals interaction, and mainly focus on two candidate Rydberg states $|r\rangle=|70S_{1/2}\rangle$ and $|r\rangle=|100S_{1/2}\rangle$, respectively, with the interaction coefficients $C_6/2\pi=858.4~{\rm GHz}~\mu {\rm m}^6$~\cite{PhysRevA.77.032723,Bernien2016} and $C_6/2\pi=56.2~{\rm THz}~\mu {\rm m}^6$~\cite{PhysRevA.77.032723,PhysRevApplied.11.044035}. The excitation of a Rydberg state in $^{87}$Rb atoms with a principal quantum number $n=70$~\cite{Bernien2016,PhysRevLett.121.123603,PhysRevLett.123.170503} or near $n=100$~\cite{PhysRevA.82.030306,PhysRevLett.104.010503} has been demonstrated. The computational states can be encoded on the hyperfine ground states $|0\rangle=|5S_{1/2}, F=1, m_{F}=1\rangle$ and $|1\rangle=|5S_{1/2}, F=2, m_{F}=2\rangle$~\cite{PhysRevLett.104.010502}. Then assisted by an intermediate state $|p\rangle=|5p_{3/2}\rangle$ or $|p\rangle=|6p_{3/2}\rangle$, the atomic transition $|1\rangle\leftrightarrow|r\rangle$ can be achieved through a two-photon process, as shown in Fig.~\ref{f1}(c)~(see the Appendix for more details).
	
In the following, we investigate the intrinsic gate error. It is noted that the intrinsic gate error here does not contain errors caused by atomic decay that will be discussed separately. In addition, other important decoherence factors, including interatomic DDF and Doppler effects, will be considered, respectively. We use $E_{\rm in}$, $E_{\rm de}$, $E_{\rm dd}$, and $E_{\rm do}$ to label the intrinsic error, atomic decay error, DDF error, and Doppler dephasing error, respectively. Each decoherence factor is addressed separately, and superscripts (1), (2), and (3) are used to label the three schemes based on the conditions $\Omega_{\rm m}=2\sqrt{3}\Omega_2$, $\Omega_{\rm m}\gg2\Omega_2$ without the LZS transition, and $\Omega_{\rm m}\gg2\Omega_2$ with the LZS transition, respectively.

\subsection{Intrinsic error}
Since the gate schemes are based on the parameter regime $V\gg\Omega_{\rm m},\Omega_2$~(and $\Omega_{\rm m}\gg2\Omega_2$ for the latter two schemes), there are intrinsic gate errors for the three schemes, similar to the blockade errors in the Rydberg-blockade-based gates,  due to the rotating-wave approximation even when any decoherence factors are not entailed, and such intrinsic errors strongly depends on the parameter setting. Here we set the interatomic distance as $d=4.8~\mu$m for $n=70$~($d=9.6~\mu$m for $n=100$), yielding the RRI strength $V/2\pi=70.18$~MHz~($V/2\pi=71.79$~MHz)~\cite{PhysRevA.77.032723,Bernien2016}. The Rabi frequency of the target-atom Rydberg pumping is set as $\Omega_2/2\pi=0.1$~MHz. In Fig.~\ref{f6}, we plot the intrinsic gate errors $E_{\rm in}=1-F$ for the three schemes changed with the evolutionary gate time, where at the ideal gate time $E_{\rm in}^{(1)}$ and $E_{\rm in}^{(2)}$ are of the same order of magnitudes, that is, $\sim10^{-6}$, while $E_{\rm in}^{(3)}\sim10^{-3}$. These intrinsic errors can be further reduced by using parameters that can ensure  more strictly the relation $V\gg\Omega_{\rm m},\Omega_2$~(and $\Omega_{\rm m}\gg2\Omega_2$ for the latter two schemes). In the later subsections, we discuss gate errors caused by several decoherence factors, and it is noted that those errors do not contain the intrinsic gate errors.

\subsection{Atomic decay}
\begin{figure}[b]
	\includegraphics[width=\linewidth]{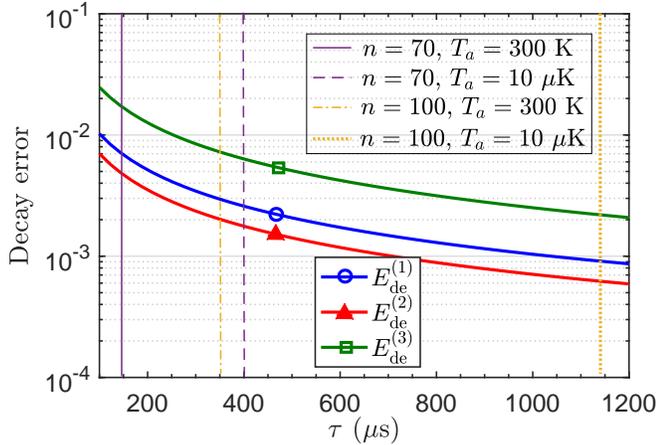}
	\caption{Decay error of the CZ gate for the three schemes with varying the Rydberg state lifetime. $n$ and $T_a$ denote the principle quantum number of the Rydberg state and the atomic temperature, respectively. Four vertical lines label the lifetimes of $|r\rangle=|70S_{1/2}\rangle$ and  $|r\rangle=|100S_{1/2}\rangle$, respectively, at the atomic temperatures $T_a=300$~K and $T_a=10~\mu$K~\cite{PhysRevApplied.11.044035,PhysRevA.79.052504}. Parameters are the same as the Fig.~\ref{f6}.}\label{f7}
\end{figure}
Atomic decay from the excited states into ground states will spoil the coherence of the desired dynamics.
In addition to the atomic decay rate, the gate fidelity reduction caused by the atomic decay is proportional to the gate time and the population of excited states.
Therefore, the scheme based on $\Omega_{\rm m}\gg\Omega_2$ without the LZS transition will suffer from the least fidelity reduction for a finite atomic decay rate, because it is of not only the shortest gate time but also the least population of excited states. The LZS-transition scheme is also of the least population of excited states, but it consumes longer gate time that is four multiple of that for the other two schemes. In order to clearly know whether or not a high-fidelity gate can be attained when the atomic decay is taken into account, we carry out numerical simulations based on the Lindblad master equation
\begin{eqnarray}\label{e11}
{\dot{\hat{\rho}}}&=&-i[\hat{H}_{\rm Full},\hat{\rho}]
-\frac{1}{2}\sum_{j=1,2}\sum_{k=0,1,g}\Big(\hat{\mathcal{L}}_{k}^{j\dag}\hat{\mathcal{L}}_{k}^j\hat{\rho}-2\hat{\mathcal{L}}_{k}^j\hat{\rho}\hat{\mathcal{L}}_{k}^{j\dag}\nonumber\\
&&+\hat{\rho}\hat{\mathcal{L}}_{k}^{j\dag}\hat{\mathcal{L}}_{k}^j\Big),
\end{eqnarray}
where $\hat{\rho}$ is the density operator and ${\dot{\hat{\rho}}}$ the time derivative of the density operator. $\hat{H}_{\rm Full}$ denotes the full Hamiltonian of the two-atom system, Eq.~(\ref{e1}) for the first and second schemes, or Eq.~(\ref{e8}) for the LZS-transition scheme. The Lindblad operator is defined by $\hat{\mathcal{L}}_{k}^j\equiv\sqrt{\gamma_k}|k\rangle_{j}\langle r|$, describing the decay of the $j$th atom from the Rydberg state $|r\rangle$ into a ground state $|k\rangle$ with a decay rate $\gamma_k$, for which an additional ground state $|g\rangle$ is introduced to denote those Zeeman magnetic sublevels out of the computational states $|0\rangle$ and $|1\rangle$. For convenience, we assume that decay rates from a Rydberg state of $^{87}$Rb atoms into the eight Zeeman ground states are identical, so $\gamma_0=\gamma_1=1/8\tau$ and $\gamma_g=3/4\tau$ with $\tau$ being the lifetime of the Rydberg state.

In Fig.~\ref{f7}, we show the decay errors of the three schemes with different Rydberg state lifetimes, and the result indicates that without the LZS transition, the condition $\Omega_{\rm m}\gg2\Omega_2$ ensures the least decay error. However, when the LZS transition is entailed, the CZ gate suffers from the most decay error, even much more than the case of $\Omega_{\rm m}=2\sqrt3\Omega_2$ which has the largest excited-state population. This is attributed to a longer gate time of the LZS-transition scheme. Nevertheless, the LZS-transition scheme may still be experimentally feasible to achieve a CZ gate with fidelity $F>0.99$ over the error correction threshold in a surface code scheme~\cite{PhysRevA.80.052312}, when the Rydberg state with a relatively high principal quantum number is employed, for example $n=100$, or when the atoms are cooled sufficiently to, for example, an atomic temperature $T_a=10~\mu$K that has been reached in experiment~\cite{PhysRevLett.121.123603,PhysRevLett.123.170503}.
\begin{figure}[htb]
	\includegraphics[width=\linewidth]{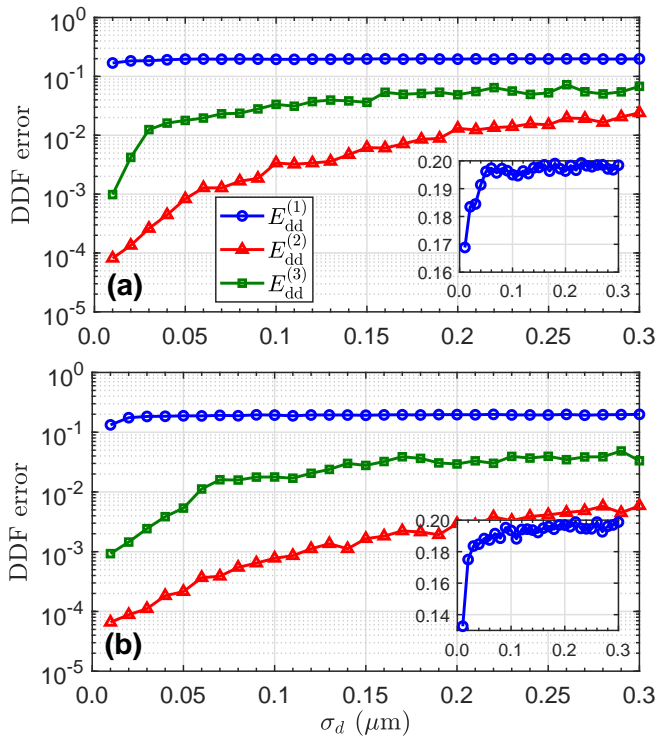}
	\caption{Dependence of the DDF error on $\sigma_d$ for the three gate schemes with (a)~$n=70$ and $d_i=4.8~\mu$m, or (b)~$n=100$ and $d_i=9.6~\mu$m. Each point denotes the average of 201 results originated from 201 samples of $d$ that are picked randomly according to a Gaussian probability distribution with the mean $d_i$ and the standard deviation $\sigma_d$. Other parameters are the same as Fig.~\ref{f6}.}\label{f8}
\end{figure}

\subsection{Interatomic dipole-dipole force}
When two adjacent atoms are both excited to the Rydberg state, the gradient of the RRI potential between the two atoms leads to a strong interatomic DDF that further induces mechanical motion and decoherence of the atoms~\cite{PhysRevLett.110.213005}, which is one of the intractable issues that make the implementation challenging of quantum gate in the RAB regime. When this interatomic DDF is taken into account, an extra term $\hat H_{\rm dd}=\frac{\partial V}{\partial d}\Big|_{d=d_i}(d-d_i)|rr\rangle\langle rr|$ should be added into the Hamiltonian in Eq.~(\ref{e1}) or Eq.~(\ref{e8})~\cite{PhysRevA.91.012337}, where $d_i=4.8~\mu$m for $n=70$ or $d_i=9.6~\mu$m for $n=100$ is the (preset) ideal interatomic distance. $d_i$ can also be regarded as the mean interatomic distance when considering that the interatomic distance $d$ is a random variable with a Gaussian probability distribution $P(d)=\exp[-{(d-d_i)^2}/{2\sigma_d^2}]/\sqrt{2\pi}\sigma_d$ with $\sigma_d$ being the standard deviation.

In Fig.~\ref{f8} we plot the dependence of the DDF error on $\sigma_d$ for the three gate schemes with $n=70$ or $n=100$, respectively. On the one hand, comparing Fig.~\ref{f8}(a) with Fig.~\ref{f8}(b), using the Rydberg state with $n=100$ can give a smaller DDF error, because the ratio $\sigma_d/d$ is smaller. On the other hand, the schemes based on the condition $\Omega_{\rm m}\gg2\Omega_2$ are always of smaller DDF errors at least one order than the first scheme, which is attributed to the suppression of excitation of the Rydberg pair state $|rr\rangle$. For even a small standard deviation of the interatomic distance, $\sigma_d=0.01~\mu$m, $E_{\rm dd}^{(1)}$ is over 0.13 that makes the CZ gate inefficient. For an experimentally accessible standard deviation of the interatomic distance, for example $\sigma_d=0.14~\mu$m reported in a recent experiment~\cite{PhysRevLett.124.033603}, the scheme based on $\Omega_{\rm m}\gg2\Omega_2$ without the LZS transition ensures a DDF error $E_{\rm dd}^{(2)}<10^{-2}$.

\subsection{Doppler effects}
Doppler effects originated from atomic motion due to finite atomic temperature cause motional dephasing of ground-Rydberg transitions, which is an important resource of technical errors~\cite{PhysRevLett.104.010502,Saffman_2011,Saffman_2016}, and some works have been contributed recently to suppressing Dopplor dephasing errors in implementing Rydberg-mediated quantum gates~\cite{PhysRevA.84.053409,PhysRevApplied.11.044035,PhysRevApplied.13.024008,sun2019doppler}. In this subsection, we study Doppler dephasing errors of the three gate schemes. In order to incur less Doppler dephasing errors, we use two counterpropagating laser beams to excite the atoms, which can reduce the effective wave vector of magnitude, and thus the standard deviation of atomic positions can be decreased effectively~\cite{Saffman_2016,PhysRevApplied.11.044035,PhysRevApplied.10.034006}. To drive Rydberg excitation through a two-photon process, we use two counterpropagating laser fields with wavelengths $\lambda_1=420$~nm and $\lambda_2=1013$~nm for using the Rydberg state $|r\rangle=|70S_{1/2}\rangle$ and the intermediate state $|p\rangle=|6P_{3/2}\rangle$, yielding an effective wave vector of magnitude $k_{\rm eff}=8.76\times10^6~{\rm m}^{-1}$~\cite{PhysRevLett.121.123603}. Due to the presence of the Doppler effects, the Rabi frequencies for the Rydberg excitation are changed, as~\cite{PhysRevA.97.053803}
\begin{equation*}
\Omega_1(t)\rightarrow\Omega_1(t)e^{i\delta_1t},\quad\Omega_2\rightarrow\Omega_2e^{i\delta_2t}.
\end{equation*}
The detunings $\delta_{1,2}$ of the excitation lasers seen by the atoms are two random variables following a Gaussian probability distribution of the mean $\delta_0=0$ and the
standard deviation $\sigma_{\delta}=k_{\rm eff}\Delta v$ where $\Delta v=\sqrt{k_{\rm B}T_a/m}$ denotes the atomic root-mean-square velocity with $k_{\rm B}$, $T_a$, and $m$ being the Boltzmann constant, atomic temperature, and atomic mass, respectively.

\begin{figure*}
	\includegraphics[width=\linewidth]{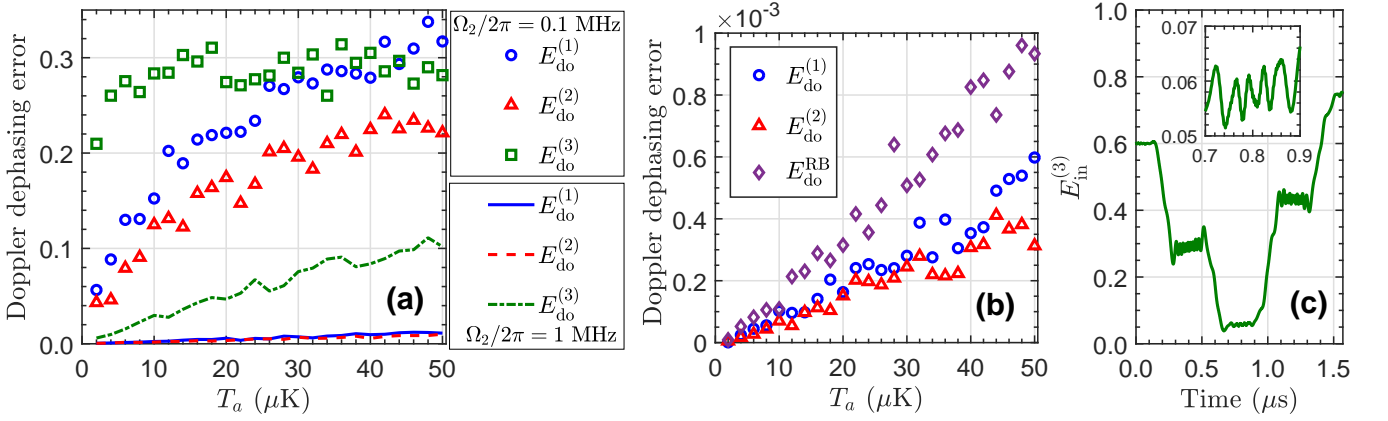}
	\caption{(a)~Doppler dephasing errors for the three CZ gate schemes with varying the atomic temperature. Two sets of parameters are used: \{$V/2\pi=70.18$~MHz~($n=70$, $d=4.8~\mu$m), $\Omega_{\rm m}/2\pi=10~$MHz, $\Omega_2/2\pi=0.1~$MHz\} and \{$V/2\pi=467$~MHz~($n=70$, $d=3.5~\mu$m), $\Omega_{\rm m}/2\pi=80~$MHz, $\Omega_2/2\pi=1~$MHz\}. (b)~Doppler dephasing errors for the first and second CZ gate schemes with varying the atomic temperature, using parameters \{$V/2\pi=467$~MHz, $\Omega_{\rm m}/2\pi=80~$MHz, $\Omega_2/2\pi=5~$MHz\}. $E_{\rm do}^{\rm RB}$ denotes the Doppler dephasing error of the standard Rydberg-blockade CZ gate in Ref.~\cite{PhysRevLett.85.2208}, using parameters \{$V/2\pi=467$~MHz, $\Omega_r/2\pi=10~$MHz\}. (c)~Time-dependence of the intrinsic error of the LZS-transition scheme, using parameters \{$V/2\pi=467$~MHz, $\Omega_{\rm m}/2\pi=80~$MHz, $\Omega_2/2\pi=5~$MHz\}. Each point in (a) and (b) denotes the average of 201 results originated from 201 pairs of $\delta_1$ and $\delta_2$ that are separately picked randomly according to a Gaussian probability distribution with the mean $\delta_0=0$ and the standard deviation $\sigma_{\delta}$.}\label{f9}
\end{figure*}
When the atomic species is determined, the Doppler dephasing errors of the three CZ gate schemes are determined by the atomic temperature. In Fig.~\ref{f9}(b), we use two sets of parameters, \{$V/2\pi=70.18$~MHz~($n=70$, $d=4.8~\mu$m), $\Omega_{\rm m}/2\pi=10~$MHz, $\Omega_2/2\pi=0.1~$MHz\} and \{$V/2\pi=467$~MHz~($n=70$, $d=3.5~\mu$m), $\Omega_{\rm m}/2\pi=80~$MHz, $\Omega_2/2\pi=1~$MHz\} that cause intrinsic gate errors with the same order of magnitudes $E_{\rm in}^{(1,2)}\sim10^{-6}$ and $E_{\rm in}^{(3)}\sim10^{-3}$, to study the effect of the atomic temperature on the Doppler dephasing errors for the three CZ gate schemes. The set of parameters with $\Omega_2/2\pi=0.1$~MHz corresponds to very high Doppler dephasing errors for all the three gate schemes even with a very low atomic temperature, because $\Omega_2/2\pi=0.1$~MHz corresponds to long Rydberg excitation durations, especially for the LZS-transition scheme. When the set of parameters with $\Omega_2/2\pi=1$~MHz is used, the Rydberg excitation duration of the each gate scheme is reduced by a order of magnitudes, and accordingly the Doppler dephasing error is slashed. For the scheme based on $\Omega_{\rm m}\gg2\Omega_2$ without the LZS transition, $E_{\rm do}^{(2)}<10^{-2}$ when $T_a<46~\mu$K.

Furthermore, when we use a larger $\Omega_2$, the Doppler dephasing errors will be suppressed to a lower level. As shown in Fig.~\ref{f9}(b), we use an alternative set of parameters \{$V/2\pi=467$~MHz, $\Omega_{\rm m}/2\pi=80~$MHz, $\Omega_2/2\pi=5~$MHz\} that gives $E_{\rm in}^{(1,2)}\sim10^{-4}$ to calculate numerically $E_{\rm do}^{(1,2)}$ with different $T_a$. When using such a set of parameters, the LZS-transition scheme is nearly inefficient because of a high intrinsic gate error $E_{\rm in}^{(3)}>5\%$~[see Fig.~\ref{f9}(c)], so $E_{\rm do}^{(3)}$ is not shown. Figure~\ref{f9}(b) shows very small Doppler dephasing errors for the first and second schemes, $E_{\rm do}^{(1)}\lesssim0.6\times10^{-3}$ and $E_{\rm do}^{(2)}\lesssim0.4\times10^{-3}$ when $T_a\in(0,50]~\mu$K. For the sake of contrast, we also calculate the Doppler dephasing error $E_{\rm do}^{\rm RB}$ of the standard Rydberg-blockade CZ gate in a three-pulse $\pi$--$2\pi$--$\pi$ scheme proposed first in Ref.~\cite{PhysRevLett.85.2208}, using a set of parameters \{$V/2\pi=467$~MHz, $\Omega_r/2\pi=10~$MHz\} with $\Omega_r$ being the Rabi frequencies of Rydberg excitation driven by the three sequential pulses, which gives an intrinsic blockade error $E_{\rm in}^{\rm RB}\sim10^{-4}$ and ensures the same gate time as the present first and second schemes shown in Fig.~\ref{f9}(b). From Fig.~\ref{f9}(b), we learn that the Doppler dephasing errors of the CZ gates in the present first and second schemes are of the same order as but lower than the standard Rydberg-blockade CZ gate.

\section{multiqubit phase gates}
\subsection{Three-qubit phase gate}
The direct implementation of multiqubit gates can significantly improve the efficiency of quantum computing in contrary to the composition with a set of single- and two-qubit gates. Therefore, the feasibility of generalizing a two-qubit gate to multiqubit cases manifests the scalability and capability of quantum computing schemes. In this section, we generalize the gate schemes to multiatom cases.

\begin{figure}[b]
\includegraphics[width=\linewidth]{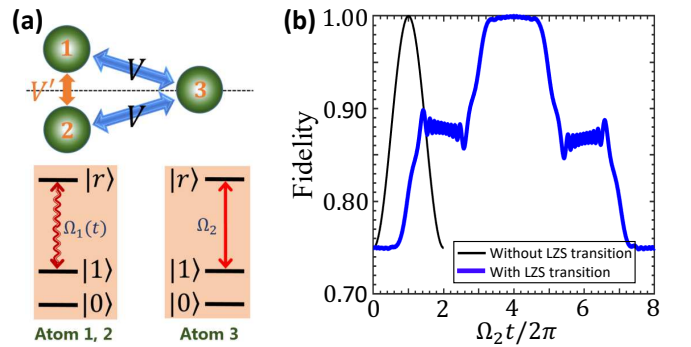}
\caption{(a)~Schematic diagram of the three-qubit phase gate. Atom~3 is located at the perpendicular bisectionplane of the line segment from Atom~1 to Atom~2. (b)~Fidelity evolution of the three-qubit phase gate based on $\Omega_{\rm m}\gg2\Omega_2$ without or with the LZS transition. $\bar\Delta=6\Omega_2$, $\Delta_0=5\Omega_2$, $\bar\omega=0.5\Omega_2$, $\Omega_{\rm m}=100\Omega_2$ and $V'=V=\omega=500\Omega_2$.}\label{f10}
\end{figure}
First of all, we introduce a three-atom system where Atom~1 and Atom~2 play the roles of control atoms driven by the amplitude-modulated fields and Atom~3 is the target atom driven by the constant-amplitude field. We specify Atom~3 located at the perpendicular bisectionplane of the line segment from Atom~1 to Atom~2 such that the RRI strength between Atom~3 and Atom~1 and that between Atom~3 and Atom~2 are identical $V=\omega$. We label the RRI strength between Atom~1 and Atom~2 with $V'$.  The schematic diagram of such a system is shown in Fig.~\ref{f10}(a), which is described by the Hamiltonian
\begin{eqnarray}\label{e12}
\hat{H}_{I3}&=&\Big(\sum_{j=1,2}\frac{\Omega_1}2|1\rangle_j\langle r|+{\rm H.c.}+V|rr\rangle_{j3}\langle rr|\Big)+\Delta|r\rangle_3\langle r|\nonumber\\
&&+\Big(\frac{\Omega_2}2|1\rangle_3\langle r|+{\rm H.c.}\Big)+V'|rr\rangle_{12}\langle rr|,
\end{eqnarray}
where $\Delta=0$~($\Delta=\Delta_0+\bar{\Delta}\cos\bar{\omega}t$) corresponds to the scheme without~(with) the LZS transition.
For convenience, we use $\Delta=0$ to interpret the construction of a three-qubit phase gate. It is easy to understand that when Atom~1 or Atom~2 is in $|0\rangle$ initially, the evolution of the other two atoms is the same as the two-atom case, which means the following transformations~(omitting the subscript of the state $|\cdot\rangle_{123}$)
\begin{eqnarray}\label{e13}
&&|000\rangle\mapsto|000\rangle,\quad|001\rangle\mapsto-|001\rangle,\quad|010\rangle\mapsto|010\rangle,\nonumber\\
&&|011\rangle\mapsto|011\rangle,\quad|1 00\rangle\mapsto|1 00\rangle,\quad|101\rangle\mapsto|101\rangle.
\end{eqnarray}
When the three atoms are in $|110\rangle$ initially, the three-atom system can be dominated by the Hamiltonian
\begin{eqnarray}\label{e14}
\hat{H}_{110}&=&\frac{\sqrt2\Omega_{\rm m}}4\Big[|110\rangle\langle \phi|\left(e^{-i\omega t}+e^{i\omega t}\right)+|\phi\rangle\langle rr0|\nonumber\\
&&\times\Big(e^{-i(\omega+V' )t}+e^{i(\omega-V' ) t}\Big)\Big]+{\rm H.c.},
\end{eqnarray}
with $|\phi\rangle\equiv(|r10\rangle+|1r0\rangle)/\sqrt2$. As long as the condition $V'=\pm2\omega$ for the two-atom two-photon process $|110\rangle\leftrightarrow|rr0\rangle$
can not be strictly satisfied, the state of the three atoms remains unchanged~\cite{WU2020126039,Wu2020OL}, that is, $|11 0\rangle\mapsto|11 0\rangle$.
Finally, if the three atoms are in $|111\rangle$ initially, after neglecting highly oscillating terms similar to the derivation of Eq.~(\ref{e4}), an effective Hamiltonian of the three-atom system can be obtained
\begin{equation}\label{e15}
\hat{H}_{111}=\frac{\Omega_2}2|111\rangle\langle 11r|+\frac{\sqrt2\Omega_{\rm m}}4|11r\rangle\langle \Phi|+{\rm H.c.},
\end{equation}
with $|\Phi\rangle\equiv(|r1r\rangle+|1rr\rangle)/\sqrt2$. When $\Omega_{\rm m}\gg\Omega_2$ is considered, the evolution from $|111\rangle$ is prohibited.

According to the process above we can achieve a three-qubit phase gate with the value of $V'$ being not close to $2V$. For testing the performance of the three-qubit phase gate, based on the Hamiltonian Eq.~(\ref{e12}) with $V'=V$ we simulate in Fig.~\ref{f10}(b) the fidelity evolution for the scheme without or with the LZS transition, for which the three-atom initial state is chosen as $\frac1{2\sqrt2}\bigotimes_{m=1}^3(|0\rangle_m+|1\rangle_m)$. Figure~\ref{f10}(b) shows that the high-fidelity three-qubit phase gate can be constructed with the same gate time as the two-atom CZ gate.

\subsection{$N$-qubit phase gate}
\begin{figure}
\includegraphics[width=\linewidth]{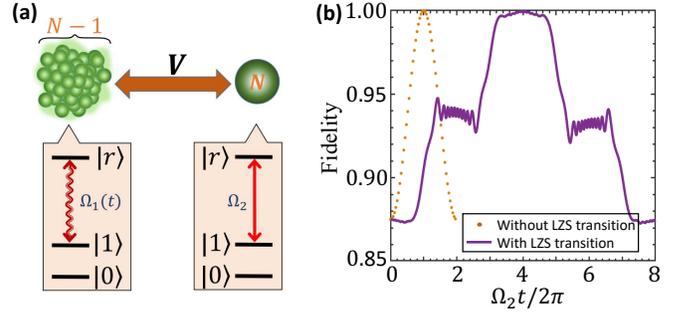}
\caption{(a)~Schematic diagram of the $N$-qubit phase gate. $(N-1)$ control atoms are set very close to each other. (b)~Fidelity evolution of the four-qubit phase gate based on $\Omega_{\rm m}\gg2\Omega_2$ without or with the LZS transition. $\bar\Delta=6\Omega_2$, $\Delta_0=5\Omega_2$, $\bar\omega=0.5\Omega_2$, $\Omega_{\rm m}=100\Omega_2$, $V_{14}=V_{24}=V_{34}=\omega=500\Omega_2$, $V_{12}=100V$, $V_{13}=200V$, and $V_{23}=500V$.}\label{f11}
\end{figure}
For $N$-qubit phase gates, we can consider that ($N-1$) control atoms are set close enough to each other to make the RRI strengths between each of them and Atom~$N$ are equivalent approximatively, as shown in Fig.~\ref{f11}(a), which can enable an $N$-qubit phase gate. When the $(N-1)$ control atoms are put within a relatively small region, the Rydberg blockade works for them and thus this ensemble of the $(N-1)$ atoms serves as a Rydberg superatom~\cite{PhysRevLett.87.037901,PhysRevLett.121.103601}. Except the collective state $\bigotimes_{j=1}^{N-1}|0\rangle_j$ where all the $(N-1)$ atoms are in $|0\rangle$, all of other collective states will make the atomic ensemble equivalent to a two-level atom, where only one of the atoms in $|1\rangle$ can be excited. Therefore, the evolution of the $N$-atom system shown in Fig.~\ref{f11}(a) is similar to that of the two-atom case. Hence at the time $T=2\pi/\Omega_2$ only the collective state $\bigotimes_{j=1}^{N-1}|0\rangle_j\otimes|1\rangle_N$ of the $N$-atom system evolves back to itself but gets a $\pi$ phase, while other collective states remain unchanged. As an example, we simulate the implementation of a four-qubit phase gate with a four-atom system governed by the Hamiltonian
\begin{eqnarray}\label{e16}
\hat{H}_{I4}&=&\Big(\sum_{j=1,2,3}\frac{\Omega_1}2|1\rangle_j\langle r|+{\rm H.c.}+V_{j4}|rr\rangle_{j4}\langle rr|\Big)\nonumber\\
&&+\Delta|r\rangle_4\langle r|+\left(\frac{\Omega_2}2|1\rangle_4\langle r|+{\rm H.c.}\right)+V_{12}|rr\rangle_{12}\langle rr|\nonumber\\
&&+V_{13}|rr\rangle_{13}\langle rr|+V_{23}|rr\rangle_{23}\langle rr|.
\end{eqnarray}
Taking an example of $V_{12}=100V$, $V_{13}=200V$, and $V_{23}=500V$, in Fig.~\ref{f11}(b) we plot the fidelity evolution of the four-qubit phase gate based on the condition $\Omega_{\rm m}\gg2\Omega_2$ without or with the LZS transition, for which the four-atom initial state is chosen as $\frac14\bigotimes_{m=1}^4(|0\rangle_m+|1\rangle_m)$. Figure~\ref{f11}(b) shows that the high-fidelity multiqubit phase gate can be attained with the same gate time as the two-atom CZ gate, and the performance is almost as fine as that of the two-atom CZ gate.

\begin{figure}[htb]
	\includegraphics[width=\linewidth]{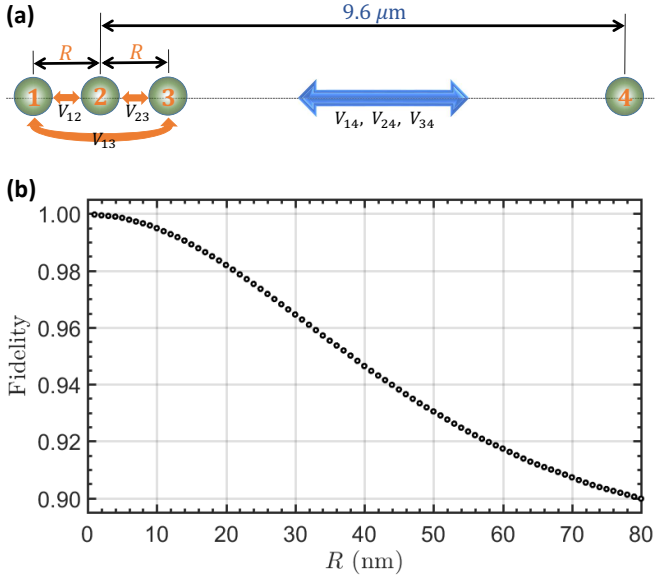}
	\caption{(a)~Schematic diagram of implementing a four-qubit phase gate. Four atoms are located inhomogeneously at a one-dimensional chain. (b)~Dependence of the gate fidelity on the radius of the ensemble of the control atoms. $|r\rangle=|100S_{1/2}\rangle$, $\Omega_{\rm m}/2\pi=10$~MHz, and $\Omega_2/2\pi=0.1$~MHz.}\label{f12}
\end{figure}

For the implementation of multiqubit phase gates shown in Fig.~\ref{f11}, the control atoms can be set as close as possible to each other so that they can be driven by solely one field~(or two fields for the two-photon process). To investigate the influence of the radius of the ensemble of the control atoms on the performance of the multiqubit phase gates, we take an example of a four-atom system shown in Fig.~\ref{f12}(a), where the four atoms are located inhomogeneously at a one-dimensional chain.
The distance between Atom~1~(Atom~3) and Atom~2 is $R$~(i.e., the radius of the ensemble of the control atoms), and that between Atom~2 and Atom~4 is $9.6~\mu$m for $|r\rangle=|100S_{1/2}\rangle$. The Hamiltonian of the four-atom system is given in Eq.~(\ref{e16}). Then in Fig.~\ref{f12}(b) using $\Omega_{\rm m}/2\pi=10$~MHz and $\Omega_{2}/2\pi=0.1$~MHz without the LZS transition, we plot the dependence of the gate fidelity at the operation time $T=10~\mu$s on the radius of the ensemble of the control atoms. With the increase of the control-atom ensemble radius the gate fidelity is reduced, because the inequality error in the RRI strengths between each control atom and the target atom increase with growing $R$. Figure~\ref{f12}(b) manifests that for achieving a high-fidelity multiqubit phase gate the radius of the ensemble of the control atoms is supposed to be less than $15~$nm.

\section{Conclusion}
We have proposed to implement resilient quantum phase gates on Rydberg atoms. The gates are based on an unconventional Rydberg anti-blockade~(RAB) regime that can provide the fast first-order dynamics. This RAB regime is induced through offsetting the Rydberg-Rydberg interaction~(RRI) by a periodical amplitude modulation of the field on the control atom. The sensitivity of the phase gates to the fluctuations in the RRI strength can be slashed through strengthening the amplitude-modulated field. Furthermore, the Landau-Zener-St\"{u}ckelberg transition is introduced on the target atom, which makes the phase gate resilient to the deviation in the gate time and the drift in the pulse amplitude. Through estimating gate errors caused by atomic decay, interatomic dipole-dipole force, and Doppler effects, it is of great potential to implement a high-fidelity quantum gate with suitable parameters under the state-of-the-art experimental conditions. Finally, we generalize the two-atom gate to $N$-atom cases, multiqubit phase gates can also be constructed in one step with the unchanged gate time.\\

\section*{Acknowledgements}
The authors acknowledge funding from the  National Natural Science Foundation of China (NSFC) (11675046, 21973023, 11804308); Program for Innovation Research of Science in Harbin Institute of Technology (A201412); Postdoctoral Scientific Research Developmental Fund of Heilongjiang Province (LBH-Q15060); and Natural Science Foundation of Henan Province under Grant No.~202300410481.

\section*{APPENDIX: Two-photon Rydberg pumping}
The two-photon process in each atom is achieved by two laser fields. For a control~(target) atom, one field is imposed for the optical excitation $|1\rangle\leftrightarrow|p\rangle$ with Rabi frequency $\Omega_{1p}$~($\Omega_{2p}$) and a red detuning $\Delta_1$~($\Delta_2$), and the other for the Rydberg excitation $|p\rangle\leftrightarrow|r\rangle$ with Rabi frequency $\Omega_{1r}$~($\Omega_{2r}$) and a blue detuning $\Delta_1$~($\Delta_2$). The field on the control atoms for $|p\rangle\leftrightarrow|r\rangle$ is modulated in amplitude $\Omega_{1r}=\tilde{\Omega}_{\rm m}\cos\omega t$, while the field on the target atom for $|p\rangle\leftrightarrow|r\rangle$ is, for only the LZS transition, modulated in frequency. Figure~\ref{f1}(c) shows the schematic diagram for the two-photon Rydberg excitations. Corresponding to the model in Fig.~\ref{f1}(a), the quantities in the Hamiltonian~(\ref{e1}) are $\Omega_{\rm m}=\tilde{\Omega}_{\rm m}\Omega_{1p}/2\Delta_1$ and $\Omega_2=\Omega_{2r}\Omega_{2p}/2\Delta_2$. If the center-to-center distance between the control and target atoms is set as $d=9.6~\mu$m~($d=4.8~\mu$m) for $|r\rangle=|100S_{1/2}\rangle$~($|r\rangle=|70S_{1/2}\rangle$) that is within the range of the van der Waals interaction, correspondingly the RRI strength $V/2\pi=71.79$~MHz~(70.18~MHz). We assume $\Omega_{1p}/2\pi=\tilde{\Omega}_{\rm m}/2\pi=200$~MHz, $\Delta_1/2\pi=2$~GHz, $\Omega_{2r}/2\pi=\Omega_{2p}/2\pi=10$~MHz, and $\Delta_2/2\pi=500$~MHz, which indicates $\Omega_{\rm m}/2\pi=10$~MHz and $\Omega_2/2\pi=0.1$~MHz. It should be also noted that in addition to the two-photon effective coupling $|1\rangle\leftrightarrow|r\rangle$ there are Stark-shift terms $\Omega_{1p}^2/4\Delta_1|1\rangle_1\langle1|$, $\tilde{\Omega}_{\rm m}^2\cos^2\omega t/4\Delta_1|r\rangle_1\langle r|$, $\Omega_{2p}^2/4\Delta_2|1\rangle_2\langle1|$, and $\Omega_{2r}^2/4\Delta_2|r\rangle_2\langle r|$. These unwanted energy shifts can be eliminated through some manners, such as auxiliary fields~\cite{Zhao2017}, detuning compensations~\cite{Han:20}, and phase corrections~\cite{Vepsaaineneaau5999}.
\bibliography{apssamp}

\begin{thebibliography}{67}%
\makeatletter
\providecommand \@ifxundefined [1]{%
 \@ifx{#1\undefined}
}%
\providecommand \@ifnum [1]{%
 \ifnum #1\expandafter \@firstoftwo
 \else \expandafter \@secondoftwo
 \fi
}%
\providecommand \@ifx [1]{%
 \ifx #1\expandafter \@firstoftwo
 \else \expandafter \@secondoftwo
 \fi
}%
\providecommand \natexlab [1]{#1}%
\providecommand \enquote  [1]{``#1''}%
\providecommand \bibnamefont  [1]{#1}%
\providecommand \bibfnamefont [1]{#1}%
\providecommand \citenamefont [1]{#1}%
\providecommand \href@noop [0]{\@secondoftwo}%
\providecommand \href [0]{\begingroup \@sanitize@url \@href}%
\providecommand \@href[1]{\@@startlink{#1}\@@href}%
\providecommand \@@href[1]{\endgroup#1\@@endlink}%
\providecommand \@sanitize@url [0]{\catcode `\\12\catcode `\$12\catcode
  `\&12\catcode `\#12\catcode `\^12\catcode `\_12\catcode `\%12\relax}%
\providecommand \@@startlink[1]{}%
\providecommand \@@endlink[0]{}%
\providecommand \url  [0]{\begingroup\@sanitize@url \@url }%
\providecommand \@url [1]{\endgroup\@href {#1}{\urlprefix }}%
\providecommand \urlprefix  [0]{URL }%
\providecommand \Eprint [0]{\href }%
\providecommand \doibase [0]{http://dx.doi.org/}%
\providecommand \selectlanguage [0]{\@gobble}%
\providecommand \bibinfo  [0]{\@secondoftwo}%
\providecommand \bibfield  [0]{\@secondoftwo}%
\providecommand \translation [1]{[#1]}%
\providecommand \BibitemOpen [0]{}%
\providecommand \bibitemStop [0]{}%
\providecommand \bibitemNoStop [0]{.\EOS\space}%
\providecommand \EOS [0]{\spacefactor3000\relax}%
\providecommand \BibitemShut  [1]{\csname bibitem#1\endcsname}%
\let\auto@bib@innerbib\@empty
%</preamble>
\bibitem [{\citenamefont {Jaksch}\ \emph {et~al.}(2000)\citenamefont {Jaksch},
  \citenamefont {Cirac}, \citenamefont {Zoller}, \citenamefont {Rolston},
  \citenamefont {C\^ot\'e},\ and\ \citenamefont {Lukin}}]{PhysRevLett.85.2208}%
  \BibitemOpen
  \bibfield  {author} {\bibinfo {author} {\bibfnamefont {D.}~\bibnamefont
  {Jaksch}}, \bibinfo {author} {\bibfnamefont {J.~I.}\ \bibnamefont {Cirac}},
  \bibinfo {author} {\bibfnamefont {P.}~\bibnamefont {Zoller}}, \bibinfo
  {author} {\bibfnamefont {S.~L.}\ \bibnamefont {Rolston}}, \bibinfo {author}
  {\bibfnamefont {R.}~\bibnamefont {C\^ot\'e}}, \ and\ \bibinfo {author}
  {\bibfnamefont {M.~D.}\ \bibnamefont {Lukin}},\ }\href {\doibase
  10.1103/PhysRevLett.85.2208} {\bibfield  {journal} {\bibinfo  {journal}
  {Phys. Rev. Lett.}\ }\textbf {\bibinfo {volume} {85}},\ \bibinfo {pages}
  {2208} (\bibinfo {year} {2000})}\BibitemShut {NoStop}%
\bibitem [{\citenamefont {Saffman}\ \emph {et~al.}(2010)\citenamefont
  {Saffman}, \citenamefont {Walker},\ and\ \citenamefont
  {M\o{}lmer}}]{Saffman2010}%
  \BibitemOpen
  \bibfield  {author} {\bibinfo {author} {\bibfnamefont {M.}~\bibnamefont
  {Saffman}}, \bibinfo {author} {\bibfnamefont {T.~G.}\ \bibnamefont {Walker}},
  \ and\ \bibinfo {author} {\bibfnamefont {K.}~\bibnamefont {M\o{}lmer}},\
  }\href {\doibase 10.1103/RevModPhys.82.2313} {\bibfield  {journal} {\bibinfo
  {journal} {Rev. Mod. Phys.}\ }\textbf {\bibinfo {volume} {82}},\ \bibinfo
  {pages} {2313} (\bibinfo {year} {2010})}\BibitemShut {NoStop}%
\bibitem [{\citenamefont {Levine}\ \emph {et~al.}(2018)\citenamefont {Levine},
  \citenamefont {Keesling}, \citenamefont {Omran}, \citenamefont {Bernien},
  \citenamefont {Schwartz}, \citenamefont {Zibrov}, \citenamefont {Endres},
  \citenamefont {Greiner}, \citenamefont {Vuleti\ifmmode~\acute{c}\else
  \'{c}\fi{}},\ and\ \citenamefont {Lukin}}]{PhysRevLett.121.123603}%
  \BibitemOpen
  \bibfield  {author} {\bibinfo {author} {\bibfnamefont {H.}~\bibnamefont
  {Levine}}, \bibinfo {author} {\bibfnamefont {A.}~\bibnamefont {Keesling}},
  \bibinfo {author} {\bibfnamefont {A.}~\bibnamefont {Omran}}, \bibinfo
  {author} {\bibfnamefont {H.}~\bibnamefont {Bernien}}, \bibinfo {author}
  {\bibfnamefont {S.}~\bibnamefont {Schwartz}}, \bibinfo {author}
  {\bibfnamefont {A.~S.}\ \bibnamefont {Zibrov}}, \bibinfo {author}
  {\bibfnamefont {M.}~\bibnamefont {Endres}}, \bibinfo {author} {\bibfnamefont
  {M.}~\bibnamefont {Greiner}}, \bibinfo {author} {\bibfnamefont
  {V.}~\bibnamefont {Vuleti\ifmmode~\acute{c}\else \'{c}\fi{}}}, \ and\
  \bibinfo {author} {\bibfnamefont {M.~D.}\ \bibnamefont {Lukin}},\ }\href
  {\doibase 10.1103/PhysRevLett.121.123603} {\bibfield  {journal} {\bibinfo
  {journal} {Phys. Rev. Lett.}\ }\textbf {\bibinfo {volume} {121}},\ \bibinfo
  {pages} {123603} (\bibinfo {year} {2018})}\BibitemShut {NoStop}%
\bibitem [{\citenamefont {Omran}\ \emph {et~al.}(2019)\citenamefont {Omran},
  \citenamefont {Levine}, \citenamefont {Keesling}, \citenamefont {Semeghini},
  \citenamefont {Wang}, \citenamefont {Ebadi}, \citenamefont {Bernien},
  \citenamefont {Zibrov}, \citenamefont {Pichler}, \citenamefont {Choi},
  \citenamefont {Cui}, \citenamefont {Rossignolo}, \citenamefont {Rembold},
  \citenamefont {Montangero}, \citenamefont {Calarco}, \citenamefont {Endres},
  \citenamefont {Greiner}, \citenamefont {Vuleti{\'c}},\ and\ \citenamefont
  {Lukin}}]{Omran570}%
  \BibitemOpen
  \bibfield  {author} {\bibinfo {author} {\bibfnamefont {A.}~\bibnamefont
  {Omran}}, \bibinfo {author} {\bibfnamefont {H.}~\bibnamefont {Levine}},
  \bibinfo {author} {\bibfnamefont {A.}~\bibnamefont {Keesling}}, \bibinfo
  {author} {\bibfnamefont {G.}~\bibnamefont {Semeghini}}, \bibinfo {author}
  {\bibfnamefont {T.~T.}\ \bibnamefont {Wang}}, \bibinfo {author}
  {\bibfnamefont {S.}~\bibnamefont {Ebadi}}, \bibinfo {author} {\bibfnamefont
  {H.}~\bibnamefont {Bernien}}, \bibinfo {author} {\bibfnamefont {A.~S.}\
  \bibnamefont {Zibrov}}, \bibinfo {author} {\bibfnamefont {H.}~\bibnamefont
  {Pichler}}, \bibinfo {author} {\bibfnamefont {S.}~\bibnamefont {Choi}},
  \bibinfo {author} {\bibfnamefont {J.}~\bibnamefont {Cui}}, \bibinfo {author}
  {\bibfnamefont {M.}~\bibnamefont {Rossignolo}}, \bibinfo {author}
  {\bibfnamefont {P.}~\bibnamefont {Rembold}}, \bibinfo {author} {\bibfnamefont
  {S.}~\bibnamefont {Montangero}}, \bibinfo {author} {\bibfnamefont
  {T.}~\bibnamefont {Calarco}}, \bibinfo {author} {\bibfnamefont
  {M.}~\bibnamefont {Endres}}, \bibinfo {author} {\bibfnamefont
  {M.}~\bibnamefont {Greiner}}, \bibinfo {author} {\bibfnamefont
  {V.}~\bibnamefont {Vuleti{\'c}}}, \ and\ \bibinfo {author} {\bibfnamefont
  {M.~D.}\ \bibnamefont {Lukin}},\ }\href {\doibase 10.1126/science.aax9743}
  {\bibfield  {journal} {\bibinfo  {journal} {Science}\ }\textbf {\bibinfo
  {volume} {365}},\ \bibinfo {pages} {570} (\bibinfo {year}
  {2019})}\BibitemShut {NoStop}%
\bibitem [{\citenamefont {Gallagher}(2005)}]{Gallagher1994}%
  \BibitemOpen
  \bibfield  {author} {\bibinfo {author} {\bibfnamefont {T.~F.}\ \bibnamefont
  {Gallagher}},\ }\href@noop {} {\emph {\bibinfo {title} {Rydberg atoms}}}\
  (\bibinfo  {publisher} {Cambridge University Press},\ \bibinfo {year}
  {2005})\BibitemShut {NoStop}%
\bibitem [{\citenamefont {Lukin}\ \emph {et~al.}(2001)\citenamefont {Lukin},
  \citenamefont {Fleischhauer}, \citenamefont {Cote}, \citenamefont {Duan},
  \citenamefont {Jaksch}, \citenamefont {Cirac},\ and\ \citenamefont
  {Zoller}}]{PhysRevLett.87.037901}%
  \BibitemOpen
  \bibfield  {author} {\bibinfo {author} {\bibfnamefont {M.~D.}\ \bibnamefont
  {Lukin}}, \bibinfo {author} {\bibfnamefont {M.}~\bibnamefont {Fleischhauer}},
  \bibinfo {author} {\bibfnamefont {R.}~\bibnamefont {Cote}}, \bibinfo {author}
  {\bibfnamefont {L.~M.}\ \bibnamefont {Duan}}, \bibinfo {author}
  {\bibfnamefont {D.}~\bibnamefont {Jaksch}}, \bibinfo {author} {\bibfnamefont
  {J.~I.}\ \bibnamefont {Cirac}}, \ and\ \bibinfo {author} {\bibfnamefont
  {P.}~\bibnamefont {Zoller}},\ }\href {\doibase 10.1103/PhysRevLett.87.037901}
  {\bibfield  {journal} {\bibinfo  {journal} {Phys. Rev. Lett.}\ }\textbf
  {\bibinfo {volume} {87}},\ \bibinfo {pages} {037901} (\bibinfo {year}
  {2001})}\BibitemShut {NoStop}%
\bibitem [{\citenamefont {Ates}\ \emph {et~al.}(2007)\citenamefont {Ates},
  \citenamefont {Pohl}, \citenamefont {Pattard},\ and\ \citenamefont
  {Rost}}]{Ates2007}%
  \BibitemOpen
  \bibfield  {author} {\bibinfo {author} {\bibfnamefont {C.}~\bibnamefont
  {Ates}}, \bibinfo {author} {\bibfnamefont {T.}~\bibnamefont {Pohl}}, \bibinfo
  {author} {\bibfnamefont {T.}~\bibnamefont {Pattard}}, \ and\ \bibinfo
  {author} {\bibfnamefont {J.~M.}\ \bibnamefont {Rost}},\ }\href {\doibase
  10.1103/PhysRevLett.98.023002} {\bibfield  {journal} {\bibinfo  {journal}
  {Phys. Rev. Lett.}\ }\textbf {\bibinfo {volume} {98}},\ \bibinfo {pages}
  {023002} (\bibinfo {year} {2007})}\BibitemShut {NoStop}%
\bibitem [{\citenamefont {Jo}\ \emph {et~al.}(2020)\citenamefont {Jo},
  \citenamefont {Song}, \citenamefont {Kim},\ and\ \citenamefont
  {Ahn}}]{PhysRevLett.124.033603}%
  \BibitemOpen
  \bibfield  {author} {\bibinfo {author} {\bibfnamefont {H.}~\bibnamefont
  {Jo}}, \bibinfo {author} {\bibfnamefont {Y.}~\bibnamefont {Song}}, \bibinfo
  {author} {\bibfnamefont {M.}~\bibnamefont {Kim}}, \ and\ \bibinfo {author}
  {\bibfnamefont {J.}~\bibnamefont {Ahn}},\ }\href {\doibase
  10.1103/PhysRevLett.124.033603} {\bibfield  {journal} {\bibinfo  {journal}
  {Phys. Rev. Lett.}\ }\textbf {\bibinfo {volume} {124}},\ \bibinfo {pages}
  {033603} (\bibinfo {year} {2020})}\BibitemShut {NoStop}%
\bibitem [{\citenamefont {Carr}\ and\ \citenamefont
  {Saffman}(2013)}]{PhysRevLett.111.033607}%
  \BibitemOpen
  \bibfield  {author} {\bibinfo {author} {\bibfnamefont {A.~W.}\ \bibnamefont
  {Carr}}\ and\ \bibinfo {author} {\bibfnamefont {M.}~\bibnamefont {Saffman}},\
  }\href {\doibase 10.1103/PhysRevLett.111.033607} {\bibfield  {journal}
  {\bibinfo  {journal} {Phys. Rev. Lett.}\ }\textbf {\bibinfo {volume} {111}},\
  \bibinfo {pages} {033607} (\bibinfo {year} {2013})}\BibitemShut {NoStop}%
\bibitem [{\citenamefont {Su}\ \emph {et~al.}(2015)\citenamefont {Su},
  \citenamefont {Guo}, \citenamefont {Wang},\ and\ \citenamefont
  {Zhang}}]{PhysRevA.92.022328}%
  \BibitemOpen
  \bibfield  {author} {\bibinfo {author} {\bibfnamefont {S.-L.}\ \bibnamefont
  {Su}}, \bibinfo {author} {\bibfnamefont {Q.}~\bibnamefont {Guo}}, \bibinfo
  {author} {\bibfnamefont {H.-F.}\ \bibnamefont {Wang}}, \ and\ \bibinfo
  {author} {\bibfnamefont {S.}~\bibnamefont {Zhang}},\ }\href {\doibase
  10.1103/PhysRevA.92.022328} {\bibfield  {journal} {\bibinfo  {journal} {Phys.
  Rev. A}\ }\textbf {\bibinfo {volume} {92}},\ \bibinfo {pages} {022328}
  (\bibinfo {year} {2015})}\BibitemShut {NoStop}%
\bibitem [{\citenamefont {Liu}\ \emph {et~al.}(2019)\citenamefont {Liu},
  \citenamefont {Song}, \citenamefont {Xue}, \citenamefont {Wang},\ and\
  \citenamefont {Yung}}]{PhysRevLett.123.100501}%
  \BibitemOpen
  \bibfield  {author} {\bibinfo {author} {\bibfnamefont {B.-J.}\ \bibnamefont
  {Liu}}, \bibinfo {author} {\bibfnamefont {X.-K.}\ \bibnamefont {Song}},
  \bibinfo {author} {\bibfnamefont {Z.-Y.}\ \bibnamefont {Xue}}, \bibinfo
  {author} {\bibfnamefont {X.}~\bibnamefont {Wang}}, \ and\ \bibinfo {author}
  {\bibfnamefont {M.-H.}\ \bibnamefont {Yung}},\ }\href {\doibase
  10.1103/PhysRevLett.123.100501} {\bibfield  {journal} {\bibinfo  {journal}
  {Phys. Rev. Lett.}\ }\textbf {\bibinfo {volume} {123}},\ \bibinfo {pages}
  {100501} (\bibinfo {year} {2019})}\BibitemShut {NoStop}%
\bibitem [{\citenamefont {Shi}(2017)}]{PhysRevApplied.7.064017}%
  \BibitemOpen
  \bibfield  {author} {\bibinfo {author} {\bibfnamefont {X.-F.}\ \bibnamefont
  {Shi}},\ }\href {\doibase 10.1103/PhysRevApplied.7.064017} {\bibfield
  {journal} {\bibinfo  {journal} {Phys. Rev. Applied}\ }\textbf {\bibinfo
  {volume} {7}},\ \bibinfo {pages} {064017} (\bibinfo {year}
  {2017})}\BibitemShut {NoStop}%
\bibitem [{\citenamefont {Shi}(2019)}]{PhysRevApplied.11.044035}%
  \BibitemOpen
  \bibfield  {author} {\bibinfo {author} {\bibfnamefont {X.-F.}\ \bibnamefont
  {Shi}},\ }\href {\doibase 10.1103/PhysRevApplied.11.044035} {\bibfield
  {journal} {\bibinfo  {journal} {Phys. Rev. Applied}\ }\textbf {\bibinfo
  {volume} {11}},\ \bibinfo {pages} {044035} (\bibinfo {year}
  {2019})}\BibitemShut {NoStop}%
\bibitem [{\citenamefont {M\o{}ller}\ \emph {et~al.}(2008)\citenamefont
  {M\o{}ller}, \citenamefont {Madsen},\ and\ \citenamefont
  {M\o{}lmer}}]{PhysRevLett.100.170504}%
  \BibitemOpen
  \bibfield  {author} {\bibinfo {author} {\bibfnamefont {D.}~\bibnamefont
  {M\o{}ller}}, \bibinfo {author} {\bibfnamefont {L.~B.}\ \bibnamefont
  {Madsen}}, \ and\ \bibinfo {author} {\bibfnamefont {K.}~\bibnamefont
  {M\o{}lmer}},\ }\href {\doibase 10.1103/PhysRevLett.100.170504} {\bibfield
  {journal} {\bibinfo  {journal} {Phys. Rev. Lett.}\ }\textbf {\bibinfo
  {volume} {100}},\ \bibinfo {pages} {170504} (\bibinfo {year}
  {2008})}\BibitemShut {NoStop}%
\bibitem [{\citenamefont {Saffman}\ \emph {et~al.}(2020)\citenamefont
  {Saffman}, \citenamefont {Beterov}, \citenamefont {Dalal}, \citenamefont
  {P\'aez},\ and\ \citenamefont {Sanders}}]{PhysRevA.101.062309}%
  \BibitemOpen
  \bibfield  {author} {\bibinfo {author} {\bibfnamefont {M.}~\bibnamefont
  {Saffman}}, \bibinfo {author} {\bibfnamefont {I.~I.}\ \bibnamefont
  {Beterov}}, \bibinfo {author} {\bibfnamefont {A.}~\bibnamefont {Dalal}},
  \bibinfo {author} {\bibfnamefont {E.~J.}\ \bibnamefont {P\'aez}}, \ and\
  \bibinfo {author} {\bibfnamefont {B.~C.}\ \bibnamefont {Sanders}},\ }\href
  {\doibase 10.1103/PhysRevA.101.062309} {\bibfield  {journal} {\bibinfo
  {journal} {Phys. Rev. A}\ }\textbf {\bibinfo {volume} {101}},\ \bibinfo
  {pages} {062309} (\bibinfo {year} {2020})}\BibitemShut {NoStop}%
\bibitem [{\citenamefont {Petrosyan}\ \emph {et~al.}(2017)\citenamefont
  {Petrosyan}, \citenamefont {Motzoi}, \citenamefont {Saffman},\ and\
  \citenamefont {M\o{}lmer}}]{PhysRevA.96.042306}%
  \BibitemOpen
  \bibfield  {author} {\bibinfo {author} {\bibfnamefont {D.}~\bibnamefont
  {Petrosyan}}, \bibinfo {author} {\bibfnamefont {F.}~\bibnamefont {Motzoi}},
  \bibinfo {author} {\bibfnamefont {M.}~\bibnamefont {Saffman}}, \ and\
  \bibinfo {author} {\bibfnamefont {K.}~\bibnamefont {M\o{}lmer}},\ }\href
  {\doibase 10.1103/PhysRevA.96.042306} {\bibfield  {journal} {\bibinfo
  {journal} {Phys. Rev. A}\ }\textbf {\bibinfo {volume} {96}},\ \bibinfo
  {pages} {042306} (\bibinfo {year} {2017})}\BibitemShut {NoStop}%
\bibitem [{\citenamefont {Beterov}\ \emph {et~al.}(2018)\citenamefont
  {Beterov}, \citenamefont {Hamzina}, \citenamefont {Yakshina}, \citenamefont
  {Tretyakov}, \citenamefont {Entin},\ and\ \citenamefont
  {Ryabtsev}}]{PhysRevA.97.032701}%
  \BibitemOpen
  \bibfield  {author} {\bibinfo {author} {\bibfnamefont {I.~I.}\ \bibnamefont
  {Beterov}}, \bibinfo {author} {\bibfnamefont {G.~N.}\ \bibnamefont
  {Hamzina}}, \bibinfo {author} {\bibfnamefont {E.~A.}\ \bibnamefont
  {Yakshina}}, \bibinfo {author} {\bibfnamefont {D.~B.}\ \bibnamefont
  {Tretyakov}}, \bibinfo {author} {\bibfnamefont {V.~M.}\ \bibnamefont
  {Entin}}, \ and\ \bibinfo {author} {\bibfnamefont {I.~I.}\ \bibnamefont
  {Ryabtsev}},\ }\href {\doibase 10.1103/PhysRevA.97.032701} {\bibfield
  {journal} {\bibinfo  {journal} {Phys. Rev. A}\ }\textbf {\bibinfo {volume}
  {97}},\ \bibinfo {pages} {032701} (\bibinfo {year} {2018})}\BibitemShut
  {NoStop}%
\bibitem [{\citenamefont {Su}\ \emph {et~al.}(2016)\citenamefont {Su},
  \citenamefont {Liang}, \citenamefont {Zhang}, \citenamefont {Wen},
  \citenamefont {Sun}, \citenamefont {Jin},\ and\ \citenamefont
  {Zhu}}]{PhysRevA.93.012306}%
  \BibitemOpen
  \bibfield  {author} {\bibinfo {author} {\bibfnamefont {S.-L.}\ \bibnamefont
  {Su}}, \bibinfo {author} {\bibfnamefont {E.}~\bibnamefont {Liang}}, \bibinfo
  {author} {\bibfnamefont {S.}~\bibnamefont {Zhang}}, \bibinfo {author}
  {\bibfnamefont {J.-J.}\ \bibnamefont {Wen}}, \bibinfo {author} {\bibfnamefont
  {L.-L.}\ \bibnamefont {Sun}}, \bibinfo {author} {\bibfnamefont
  {Z.}~\bibnamefont {Jin}}, \ and\ \bibinfo {author} {\bibfnamefont {A.-D.}\
  \bibnamefont {Zhu}},\ }\href {\doibase 10.1103/PhysRevA.93.012306} {\bibfield
   {journal} {\bibinfo  {journal} {Phys. Rev. A}\ }\textbf {\bibinfo {volume}
  {93}},\ \bibinfo {pages} {012306} (\bibinfo {year} {2016})}\BibitemShut
  {NoStop}%
\bibitem [{\citenamefont {Shao}\ \emph {et~al.}(2017)\citenamefont {Shao},
  \citenamefont {Li}, \citenamefont {Ji}, \citenamefont {Wu},\ and\
  \citenamefont {Yi}}]{PhysRevA.96.012328}%
  \BibitemOpen
  \bibfield  {author} {\bibinfo {author} {\bibfnamefont {X.~Q.}\ \bibnamefont
  {Shao}}, \bibinfo {author} {\bibfnamefont {D.~X.}\ \bibnamefont {Li}},
  \bibinfo {author} {\bibfnamefont {Y.~Q.}\ \bibnamefont {Ji}}, \bibinfo
  {author} {\bibfnamefont {J.~H.}\ \bibnamefont {Wu}}, \ and\ \bibinfo {author}
  {\bibfnamefont {X.~X.}\ \bibnamefont {Yi}},\ }\href {\doibase
  10.1103/PhysRevA.96.012328} {\bibfield  {journal} {\bibinfo  {journal} {Phys.
  Rev. A}\ }\textbf {\bibinfo {volume} {96}},\ \bibinfo {pages} {012328}
  (\bibinfo {year} {2017})}\BibitemShut {NoStop}%
\bibitem [{\citenamefont {Su}\ \emph {et~al.}(2017{\natexlab{a}})\citenamefont
  {Su}, \citenamefont {Gao}, \citenamefont {Liang},\ and\ \citenamefont
  {Zhang}}]{PhysRevA.95.022319}%
  \BibitemOpen
  \bibfield  {author} {\bibinfo {author} {\bibfnamefont {S.-L.}\ \bibnamefont
  {Su}}, \bibinfo {author} {\bibfnamefont {Y.}~\bibnamefont {Gao}}, \bibinfo
  {author} {\bibfnamefont {E.}~\bibnamefont {Liang}}, \ and\ \bibinfo {author}
  {\bibfnamefont {S.}~\bibnamefont {Zhang}},\ }\href {\doibase
  10.1103/PhysRevA.95.022319} {\bibfield  {journal} {\bibinfo  {journal} {Phys.
  Rev. A}\ }\textbf {\bibinfo {volume} {95}},\ \bibinfo {pages} {022319}
  (\bibinfo {year} {2017}{\natexlab{a}})}\BibitemShut {NoStop}%
\bibitem [{\citenamefont {Wu}\ \emph {et~al.}(2020{\natexlab{a}})\citenamefont
  {Wu}, \citenamefont {Song},\ and\ \citenamefont {Su}}]{WU2020126039}%
  \BibitemOpen
  \bibfield  {author} {\bibinfo {author} {\bibfnamefont {J.-L.}\ \bibnamefont
  {Wu}}, \bibinfo {author} {\bibfnamefont {J.}~\bibnamefont {Song}}, \ and\
  \bibinfo {author} {\bibfnamefont {S.-L.}\ \bibnamefont {Su}},\ }\href
  {\doibase https://doi.org/10.1016/j.physleta.2019.126039} {\bibfield
  {journal} {\bibinfo  {journal} {Phys. Lett. A}\ }\textbf {\bibinfo {volume}
  {384}},\ \bibinfo {pages} {126039} (\bibinfo {year}
  {2020}{\natexlab{a}})}\BibitemShut {NoStop}%
\bibitem [{\citenamefont {Su}\ \emph {et~al.}(2017{\natexlab{b}})\citenamefont
  {Su}, \citenamefont {Tian}, \citenamefont {Shen}, \citenamefont {Zang},
  \citenamefont {Liang},\ and\ \citenamefont {Zhang}}]{PhysRevA.96.042335}%
  \BibitemOpen
  \bibfield  {author} {\bibinfo {author} {\bibfnamefont {S.-L.}\ \bibnamefont
  {Su}}, \bibinfo {author} {\bibfnamefont {Y.}~\bibnamefont {Tian}}, \bibinfo
  {author} {\bibfnamefont {H.~Z.}\ \bibnamefont {Shen}}, \bibinfo {author}
  {\bibfnamefont {H.}~\bibnamefont {Zang}}, \bibinfo {author} {\bibfnamefont
  {E.}~\bibnamefont {Liang}}, \ and\ \bibinfo {author} {\bibfnamefont
  {S.}~\bibnamefont {Zhang}},\ }\href {\doibase 10.1103/PhysRevA.96.042335}
  {\bibfield  {journal} {\bibinfo  {journal} {Phys. Rev. A}\ }\textbf {\bibinfo
  {volume} {96}},\ \bibinfo {pages} {042335} (\bibinfo {year}
  {2017}{\natexlab{b}})}\BibitemShut {NoStop}%
\bibitem [{\citenamefont {Wu}\ \emph {et~al.}(2020{\natexlab{b}})\citenamefont
  {Wu}, \citenamefont {Su}, \citenamefont {Wang}, \citenamefont {Song},
  \citenamefont {Xia},\ and\ \citenamefont {Jiang}}]{Wu2020OL}%
  \BibitemOpen
  \bibfield  {author} {\bibinfo {author} {\bibfnamefont {J.-L.}\ \bibnamefont
  {Wu}}, \bibinfo {author} {\bibfnamefont {S.-L.}\ \bibnamefont {Su}}, \bibinfo
  {author} {\bibfnamefont {Y.}~\bibnamefont {Wang}}, \bibinfo {author}
  {\bibfnamefont {J.}~\bibnamefont {Song}}, \bibinfo {author} {\bibfnamefont
  {Y.}~\bibnamefont {Xia}}, \ and\ \bibinfo {author} {\bibfnamefont
  {Y.}~\bibnamefont {Jiang}},\ }\href {\doibase 10.1364/OL.386765} {\bibfield
  {journal} {\bibinfo  {journal} {Opt. Lett.}\ }\textbf {\bibinfo {volume}
  {45}},\ \bibinfo {pages} {1200} (\bibinfo {year}
  {2020}{\natexlab{b}})}\BibitemShut {NoStop}%
\bibitem [{\citenamefont {Su}\ \emph {et~al.}(2020)\citenamefont {Su},
  \citenamefont {Guo}, \citenamefont {Tian}, \citenamefont {Zhu}, \citenamefont
  {Yan}, \citenamefont {Liang},\ and\ \citenamefont
  {Feng}}]{PhysRevA.101.012347}%
  \BibitemOpen
  \bibfield  {author} {\bibinfo {author} {\bibfnamefont {S.-L.}\ \bibnamefont
  {Su}}, \bibinfo {author} {\bibfnamefont {F.-Q.}\ \bibnamefont {Guo}},
  \bibinfo {author} {\bibfnamefont {L.}~\bibnamefont {Tian}}, \bibinfo {author}
  {\bibfnamefont {X.-Y.}\ \bibnamefont {Zhu}}, \bibinfo {author} {\bibfnamefont
  {L.-L.}\ \bibnamefont {Yan}}, \bibinfo {author} {\bibfnamefont {E.-J.}\
  \bibnamefont {Liang}}, \ and\ \bibinfo {author} {\bibfnamefont
  {M.}~\bibnamefont {Feng}},\ }\href {\doibase 10.1103/PhysRevA.101.012347}
  {\bibfield  {journal} {\bibinfo  {journal} {Phys. Rev. A}\ }\textbf {\bibinfo
  {volume} {101}},\ \bibinfo {pages} {012347} (\bibinfo {year}
  {2020})}\BibitemShut {NoStop}%
\bibitem [{\citenamefont {Zhu}\ \emph {et~al.}(2019)\citenamefont {Zhu},
  \citenamefont {Liang},\ and\ \citenamefont {Su}}]{Zhu:19}%
  \BibitemOpen
  \bibfield  {author} {\bibinfo {author} {\bibfnamefont {X.-Y.}\ \bibnamefont
  {Zhu}}, \bibinfo {author} {\bibfnamefont {E.}~\bibnamefont {Liang}}, \ and\
  \bibinfo {author} {\bibfnamefont {S.-L.}\ \bibnamefont {Su}},\ }\href
  {\doibase 10.1364/JOSAB.36.001937} {\bibfield  {journal} {\bibinfo  {journal}
  {J. Opt. Soc. Am. B}\ }\textbf {\bibinfo {volume} {36}},\ \bibinfo {pages}
  {1937} (\bibinfo {year} {2019})}\BibitemShut {NoStop}%
\bibitem [{\citenamefont {Li}\ \emph {et~al.}(2013)\citenamefont {Li},
  \citenamefont {Ates},\ and\ \citenamefont
  {Lesanovsky}}]{PhysRevLett.110.213005}%
  \BibitemOpen
  \bibfield  {author} {\bibinfo {author} {\bibfnamefont {W.}~\bibnamefont
  {Li}}, \bibinfo {author} {\bibfnamefont {C.}~\bibnamefont {Ates}}, \ and\
  \bibinfo {author} {\bibfnamefont {I.}~\bibnamefont {Lesanovsky}},\ }\href
  {\doibase 10.1103/PhysRevLett.110.213005} {\bibfield  {journal} {\bibinfo
  {journal} {Phys. Rev. Lett.}\ }\textbf {\bibinfo {volume} {110}},\ \bibinfo
  {pages} {213005} (\bibinfo {year} {2013})}\BibitemShut {NoStop}%
\bibitem [{\citenamefont {Basak}\ \emph {et~al.}(2018)\citenamefont {Basak},
  \citenamefont {Chougale},\ and\ \citenamefont
  {Nath}}]{PhysRevLett.120.123204}%
  \BibitemOpen
  \bibfield  {author} {\bibinfo {author} {\bibfnamefont {S.}~\bibnamefont
  {Basak}}, \bibinfo {author} {\bibfnamefont {Y.}~\bibnamefont {Chougale}}, \
  and\ \bibinfo {author} {\bibfnamefont {R.}~\bibnamefont {Nath}},\ }\href
  {\doibase 10.1103/PhysRevLett.120.123204} {\bibfield  {journal} {\bibinfo
  {journal} {Phys. Rev. Lett.}\ }\textbf {\bibinfo {volume} {120}},\ \bibinfo
  {pages} {123204} (\bibinfo {year} {2018})}\BibitemShut {NoStop}%
\bibitem [{\citenamefont {Huang}\ \emph {et~al.}(2018)\citenamefont {Huang},
  \citenamefont {Ding}, \citenamefont {Hu}, \citenamefont {Shen}, \citenamefont
  {Li}, \citenamefont {Wu},\ and\ \citenamefont {Zheng}}]{PhysRevA.98.052324}%
  \BibitemOpen
  \bibfield  {author} {\bibinfo {author} {\bibfnamefont {X.-R.}\ \bibnamefont
  {Huang}}, \bibinfo {author} {\bibfnamefont {Z.-X.}\ \bibnamefont {Ding}},
  \bibinfo {author} {\bibfnamefont {C.-S.}\ \bibnamefont {Hu}}, \bibinfo
  {author} {\bibfnamefont {L.-T.}\ \bibnamefont {Shen}}, \bibinfo {author}
  {\bibfnamefont {W.}~\bibnamefont {Li}}, \bibinfo {author} {\bibfnamefont
  {H.}~\bibnamefont {Wu}}, \ and\ \bibinfo {author} {\bibfnamefont {S.-B.}\
  \bibnamefont {Zheng}},\ }\href {\doibase 10.1103/PhysRevA.98.052324}
  {\bibfield  {journal} {\bibinfo  {journal} {Phys. Rev. A}\ }\textbf {\bibinfo
  {volume} {98}},\ \bibinfo {pages} {052324} (\bibinfo {year}
  {2018})}\BibitemShut {NoStop}%
\bibitem [{\citenamefont {Li}\ \emph {et~al.}(2020)\citenamefont {Li},
  \citenamefont {Yu}, \citenamefont {Su},\ and\ \citenamefont
  {Qian}}]{li2019periodicallydriven}%
  \BibitemOpen
  \bibfield  {author} {\bibinfo {author} {\bibfnamefont {R.}~\bibnamefont
  {Li}}, \bibinfo {author} {\bibfnamefont {D.}~\bibnamefont {Yu}}, \bibinfo
  {author} {\bibfnamefont {S.-L.}\ \bibnamefont {Su}}, \ and\ \bibinfo {author}
  {\bibfnamefont {J.}~\bibnamefont {Qian}},\ }\href {\doibase
  10.1103/PhysRevA.101.042328} {\bibfield  {journal} {\bibinfo  {journal}
  {Phys. Rev. A}\ }\textbf {\bibinfo {volume} {101}},\ \bibinfo {pages}
  {042328} (\bibinfo {year} {2020})}\BibitemShut {NoStop}%
\bibitem [{\citenamefont {Niranjan}\ \emph {et~al.}(2020)\citenamefont
  {Niranjan}, \citenamefont {Li},\ and\ \citenamefont
  {Nath}}]{niranjan2020landauzener}%
  \BibitemOpen
  \bibfield  {author} {\bibinfo {author} {\bibfnamefont {A.}~\bibnamefont
  {Niranjan}}, \bibinfo {author} {\bibfnamefont {W.}~\bibnamefont {Li}}, \ and\
  \bibinfo {author} {\bibfnamefont {R.}~\bibnamefont {Nath}},\ }\href {\doibase
  10.1103/PhysRevA.101.063415} {\bibfield  {journal} {\bibinfo  {journal}
  {Phys. Rev. A}\ }\textbf {\bibinfo {volume} {101}},\ \bibinfo {pages}
  {063415} (\bibinfo {year} {2020})}\BibitemShut {NoStop}%
\bibitem [{\citenamefont {Mallavarapu}\ \emph {et~al.}(2020)\citenamefont
  {Mallavarapu}, \citenamefont {Niranjan}, \citenamefont {Li}, \citenamefont
  {Wüster},\ and\ \citenamefont {Nath}}]{Mallavarapu2020}%
  \BibitemOpen
  \bibfield  {author} {\bibinfo {author} {\bibfnamefont {S.~K.}\ \bibnamefont
  {Mallavarapu}}, \bibinfo {author} {\bibfnamefont {A.}~\bibnamefont
  {Niranjan}}, \bibinfo {author} {\bibfnamefont {W.}~\bibnamefont {Li}},
  \bibinfo {author} {\bibfnamefont {S.}~\bibnamefont {Wüster}}, \ and\
  \bibinfo {author} {\bibfnamefont {R.}~\bibnamefont {Nath}},\ }\href@noop {}
  {\enquote {\bibinfo {title} {Population trapping in a pair of periodically
  driven rydberg atoms},}\ } (\bibinfo {year} {2020}),\ \Eprint
  {http://arxiv.org/abs/2009.10028} {arXiv:2009.10028} \BibitemShut {NoStop}%
\bibitem [{\citenamefont {van Ditzhuijzen}\ \emph {et~al.}(2009)\citenamefont
  {van Ditzhuijzen}, \citenamefont {Tauschinsky},\ and\ \citenamefont {van
  Linden van~den Heuvell}}]{PhysRevA.80.063407}%
  \BibitemOpen
  \bibfield  {author} {\bibinfo {author} {\bibfnamefont {C.~S.~E.}\
  \bibnamefont {van Ditzhuijzen}}, \bibinfo {author} {\bibfnamefont
  {A.}~\bibnamefont {Tauschinsky}}, \ and\ \bibinfo {author} {\bibfnamefont
  {H.~B.}\ \bibnamefont {van Linden van~den Heuvell}},\ }\href {\doibase
  10.1103/PhysRevA.80.063407} {\bibfield  {journal} {\bibinfo  {journal} {Phys.
  Rev. A}\ }\textbf {\bibinfo {volume} {80}},\ \bibinfo {pages} {063407}
  (\bibinfo {year} {2009})}\BibitemShut {NoStop}%
\bibitem [{\citenamefont {Shevchenko}\ \emph {et~al.}(2010)\citenamefont
  {Shevchenko}, \citenamefont {Ashhab},\ and\ \citenamefont
  {Nori}}]{SHEVCHENKO20101}%
  \BibitemOpen
  \bibfield  {author} {\bibinfo {author} {\bibfnamefont {S.}~\bibnamefont
  {Shevchenko}}, \bibinfo {author} {\bibfnamefont {S.}~\bibnamefont {Ashhab}},
  \ and\ \bibinfo {author} {\bibfnamefont {F.}~\bibnamefont {Nori}},\ }\href
  {\doibase 10.1016/j.physrep.2010.03.002} {\bibfield  {journal} {\bibinfo
  {journal} {Phys. Rep.}\ }\textbf {\bibinfo {volume} {492}},\ \bibinfo {pages}
  {1} (\bibinfo {year} {2010})}\BibitemShut {NoStop}%
\bibitem [{\citenamefont {Su}\ \emph {et~al.}(2018)\citenamefont {Su},
  \citenamefont {Shen}, \citenamefont {Liang},\ and\ \citenamefont
  {Zhang}}]{PhysRevA.98.032306}%
  \BibitemOpen
  \bibfield  {author} {\bibinfo {author} {\bibfnamefont {S.~L.}\ \bibnamefont
  {Su}}, \bibinfo {author} {\bibfnamefont {H.~Z.}\ \bibnamefont {Shen}},
  \bibinfo {author} {\bibfnamefont {E.}~\bibnamefont {Liang}}, \ and\ \bibinfo
  {author} {\bibfnamefont {S.}~\bibnamefont {Zhang}},\ }\href {\doibase
  10.1103/PhysRevA.98.032306} {\bibfield  {journal} {\bibinfo  {journal} {Phys.
  Rev. A}\ }\textbf {\bibinfo {volume} {98}},\ \bibinfo {pages} {032306}
  (\bibinfo {year} {2018})}\BibitemShut {NoStop}%
\bibitem [{\citenamefont {Xing}\ \emph {et~al.}(2020)\citenamefont {Xing},
  \citenamefont {Wu},\ and\ \citenamefont {Xu}}]{PhysRevA.101.012306}%
  \BibitemOpen
  \bibfield  {author} {\bibinfo {author} {\bibfnamefont {T.~H.}\ \bibnamefont
  {Xing}}, \bibinfo {author} {\bibfnamefont {X.}~\bibnamefont {Wu}}, \ and\
  \bibinfo {author} {\bibfnamefont {G.~F.}\ \bibnamefont {Xu}},\ }\href
  {\doibase 10.1103/PhysRevA.101.012306} {\bibfield  {journal} {\bibinfo
  {journal} {Phys. Rev. A}\ }\textbf {\bibinfo {volume} {101}},\ \bibinfo
  {pages} {012306} (\bibinfo {year} {2020})}\BibitemShut {NoStop}%
\bibitem [{\citenamefont {Shi}(2018{\natexlab{a}})}]{PhysRevApplied.9.051001}%
  \BibitemOpen
  \bibfield  {author} {\bibinfo {author} {\bibfnamefont {X.-F.}\ \bibnamefont
  {Shi}},\ }\href {\doibase 10.1103/PhysRevApplied.9.051001} {\bibfield
  {journal} {\bibinfo  {journal} {Phys. Rev. Applied}\ }\textbf {\bibinfo
  {volume} {9}},\ \bibinfo {pages} {051001} (\bibinfo {year}
  {2018}{\natexlab{a}})}\BibitemShut {NoStop}%
\bibitem [{\citenamefont {Shen}\ \emph {et~al.}(2019)\citenamefont {Shen},
  \citenamefont {Wu}, \citenamefont {Su},\ and\ \citenamefont
  {Liang}}]{Shen:19}%
  \BibitemOpen
  \bibfield  {author} {\bibinfo {author} {\bibfnamefont {C.-P.}\ \bibnamefont
  {Shen}}, \bibinfo {author} {\bibfnamefont {J.-L.}\ \bibnamefont {Wu}},
  \bibinfo {author} {\bibfnamefont {S.-L.}\ \bibnamefont {Su}}, \ and\ \bibinfo
  {author} {\bibfnamefont {E.}~\bibnamefont {Liang}},\ }\href {\doibase
  10.1364/OL.44.002036} {\bibfield  {journal} {\bibinfo  {journal} {Opt.
  Lett.}\ }\textbf {\bibinfo {volume} {44}},\ \bibinfo {pages} {2036} (\bibinfo
  {year} {2019})}\BibitemShut {NoStop}%
\bibitem [{\citenamefont {Liao}\ \emph {et~al.}(2019)\citenamefont {Liao},
  \citenamefont {Liu}, \citenamefont {Li},\ and\ \citenamefont {Du}}]{Liao:19}%
  \BibitemOpen
  \bibfield  {author} {\bibinfo {author} {\bibfnamefont {K.-Y.}\ \bibnamefont
  {Liao}}, \bibinfo {author} {\bibfnamefont {X.-H.}\ \bibnamefont {Liu}},
  \bibinfo {author} {\bibfnamefont {Z.}~\bibnamefont {Li}}, \ and\ \bibinfo
  {author} {\bibfnamefont {Y.-X.}\ \bibnamefont {Du}},\ }\href {\doibase
  10.1364/OL.44.004801} {\bibfield  {journal} {\bibinfo  {journal} {Opt.
  Lett.}\ }\textbf {\bibinfo {volume} {44}},\ \bibinfo {pages} {4801} (\bibinfo
  {year} {2019})}\BibitemShut {NoStop}%
\bibitem [{\citenamefont {Liu}\ \emph {et~al.}(2020)\citenamefont {Liu},
  \citenamefont {Su},\ and\ \citenamefont {Yung}}]{Liu2020}%
  \BibitemOpen
  \bibfield  {author} {\bibinfo {author} {\bibfnamefont {B.-J.}\ \bibnamefont
  {Liu}}, \bibinfo {author} {\bibfnamefont {S.-L.}\ \bibnamefont {Su}}, \ and\
  \bibinfo {author} {\bibfnamefont {M.-H.}\ \bibnamefont {Yung}},\ }\href
  {\doibase 10.1103/PhysRevResearch.2.043130} {\bibfield  {journal} {\bibinfo
  {journal} {Phys. Rev. Research}\ }\textbf {\bibinfo {volume} {2}},\ \bibinfo
  {pages} {043130} (\bibinfo {year} {2020})}\BibitemShut {NoStop}%
\bibitem [{\citenamefont {M\"uller}\ \emph {et~al.}(2014)\citenamefont
  {M\"uller}, \citenamefont {Murphy}, \citenamefont {Montangero}, \citenamefont
  {Calarco}, \citenamefont {Grangier},\ and\ \citenamefont
  {Browaeys}}]{PhysRevA.89.032334}%
  \BibitemOpen
  \bibfield  {author} {\bibinfo {author} {\bibfnamefont {M.~M.}\ \bibnamefont
  {M\"uller}}, \bibinfo {author} {\bibfnamefont {M.}~\bibnamefont {Murphy}},
  \bibinfo {author} {\bibfnamefont {S.}~\bibnamefont {Montangero}}, \bibinfo
  {author} {\bibfnamefont {T.}~\bibnamefont {Calarco}}, \bibinfo {author}
  {\bibfnamefont {P.}~\bibnamefont {Grangier}}, \ and\ \bibinfo {author}
  {\bibfnamefont {A.}~\bibnamefont {Browaeys}},\ }\href {\doibase
  10.1103/PhysRevA.89.032334} {\bibfield  {journal} {\bibinfo  {journal} {Phys.
  Rev. A}\ }\textbf {\bibinfo {volume} {89}},\ \bibinfo {pages} {032334}
  (\bibinfo {year} {2014})}\BibitemShut {NoStop}%
\bibitem [{\citenamefont {Mitra}\ \emph {et~al.}(2020)\citenamefont {Mitra},
  \citenamefont {Martin}, \citenamefont {Biedermann}, \citenamefont {Marino},
  \citenamefont {Poggi},\ and\ \citenamefont {Deutsch}}]{PhysRevA.101.030301}%
  \BibitemOpen
  \bibfield  {author} {\bibinfo {author} {\bibfnamefont {A.}~\bibnamefont
  {Mitra}}, \bibinfo {author} {\bibfnamefont {M.~J.}\ \bibnamefont {Martin}},
  \bibinfo {author} {\bibfnamefont {G.~W.}\ \bibnamefont {Biedermann}},
  \bibinfo {author} {\bibfnamefont {A.~M.}\ \bibnamefont {Marino}}, \bibinfo
  {author} {\bibfnamefont {P.~M.}\ \bibnamefont {Poggi}}, \ and\ \bibinfo
  {author} {\bibfnamefont {I.~H.}\ \bibnamefont {Deutsch}},\ }\href {\doibase
  10.1103/PhysRevA.101.030301} {\bibfield  {journal} {\bibinfo  {journal}
  {Phys. Rev. A}\ }\textbf {\bibinfo {volume} {101}},\ \bibinfo {pages}
  {030301(R)} (\bibinfo {year} {2020})}\BibitemShut {NoStop}%
\bibitem [{\citenamefont {Levine}\ \emph {et~al.}(2019)\citenamefont {Levine},
  \citenamefont {Keesling}, \citenamefont {Semeghini}, \citenamefont {Omran},
  \citenamefont {Wang}, \citenamefont {Ebadi}, \citenamefont {Bernien},
  \citenamefont {Greiner}, \citenamefont {Vuleti\ifmmode~\acute{c}\else
  \'{c}\fi{}}, \citenamefont {Pichler},\ and\ \citenamefont
  {Lukin}}]{PhysRevLett.123.170503}%
  \BibitemOpen
  \bibfield  {author} {\bibinfo {author} {\bibfnamefont {H.}~\bibnamefont
  {Levine}}, \bibinfo {author} {\bibfnamefont {A.}~\bibnamefont {Keesling}},
  \bibinfo {author} {\bibfnamefont {G.}~\bibnamefont {Semeghini}}, \bibinfo
  {author} {\bibfnamefont {A.}~\bibnamefont {Omran}}, \bibinfo {author}
  {\bibfnamefont {T.~T.}\ \bibnamefont {Wang}}, \bibinfo {author}
  {\bibfnamefont {S.}~\bibnamefont {Ebadi}}, \bibinfo {author} {\bibfnamefont
  {H.}~\bibnamefont {Bernien}}, \bibinfo {author} {\bibfnamefont
  {M.}~\bibnamefont {Greiner}}, \bibinfo {author} {\bibfnamefont
  {V.}~\bibnamefont {Vuleti\ifmmode~\acute{c}\else \'{c}\fi{}}}, \bibinfo
  {author} {\bibfnamefont {H.}~\bibnamefont {Pichler}}, \ and\ \bibinfo
  {author} {\bibfnamefont {M.~D.}\ \bibnamefont {Lukin}},\ }\href {\doibase
  10.1103/PhysRevLett.123.170503} {\bibfield  {journal} {\bibinfo  {journal}
  {Phys. Rev. Lett.}\ }\textbf {\bibinfo {volume} {123}},\ \bibinfo {pages}
  {170503} (\bibinfo {year} {2019})}\BibitemShut {NoStop}%
\bibitem [{\citenamefont {Yu}\ \emph {et~al.}(2020)\citenamefont {Yu},
  \citenamefont {Gao}, \citenamefont {Zhang}, \citenamefont {Liu},\ and\
  \citenamefont {Qian}}]{yu2020scalability}%
  \BibitemOpen
  \bibfield  {author} {\bibinfo {author} {\bibfnamefont {D.}~\bibnamefont
  {Yu}}, \bibinfo {author} {\bibfnamefont {Y.}~\bibnamefont {Gao}}, \bibinfo
  {author} {\bibfnamefont {W.}~\bibnamefont {Zhang}}, \bibinfo {author}
  {\bibfnamefont {J.}~\bibnamefont {Liu}}, \ and\ \bibinfo {author}
  {\bibfnamefont {J.}~\bibnamefont {Qian}},\ }\href@noop {} {\enquote {\bibinfo
  {title} {Scalability and high-efficiency of an $(n+1)$-qubit toffoli gate
  sphere via blockaded rydberg atoms},}\ } (\bibinfo {year} {2020}),\ \Eprint
  {http://arxiv.org/abs/2001.04599} {arXiv:2001.04599} \BibitemShut {NoStop}%
\bibitem [{\citenamefont {Dugan}\ \emph {et~al.}(1997)\citenamefont {Dugan},
  \citenamefont {Tull},\ and\ \citenamefont {Warren}}]{dugan1997high}%
  \BibitemOpen
  \bibfield  {author} {\bibinfo {author} {\bibfnamefont {M.~A.}\ \bibnamefont
  {Dugan}}, \bibinfo {author} {\bibfnamefont {J.~X.}\ \bibnamefont {Tull}}, \
  and\ \bibinfo {author} {\bibfnamefont {W.~S.}\ \bibnamefont {Warren}},\
  }\href {\doibase 10.1364/JOSAB.14.002348} {\bibfield  {journal} {\bibinfo
  {journal} {J. Opt. Soc. Am. B}\ }\textbf {\bibinfo {volume} {14}},\ \bibinfo
  {pages} {2348} (\bibinfo {year} {1997})}\BibitemShut {NoStop}%
\bibitem [{\citenamefont {Ashhab}\ \emph {et~al.}(2007)\citenamefont {Ashhab},
  \citenamefont {Johansson}, \citenamefont {Zagoskin},\ and\ \citenamefont
  {Nori}}]{PhysRevA.75.063414}%
  \BibitemOpen
  \bibfield  {author} {\bibinfo {author} {\bibfnamefont {S.}~\bibnamefont
  {Ashhab}}, \bibinfo {author} {\bibfnamefont {J.~R.}\ \bibnamefont
  {Johansson}}, \bibinfo {author} {\bibfnamefont {A.~M.}\ \bibnamefont
  {Zagoskin}}, \ and\ \bibinfo {author} {\bibfnamefont {F.}~\bibnamefont
  {Nori}},\ }\href {\doibase 10.1103/PhysRevA.75.063414} {\bibfield  {journal}
  {\bibinfo  {journal} {Phys. Rev. A}\ }\textbf {\bibinfo {volume} {75}},\
  \bibinfo {pages} {063414} (\bibinfo {year} {2007})}\BibitemShut {NoStop}%
\bibitem [{\citenamefont {Graham}\ \emph {et~al.}(2019)\citenamefont {Graham},
  \citenamefont {Kwon}, \citenamefont {Grinkemeyer}, \citenamefont {Marra},
  \citenamefont {Jiang}, \citenamefont {Lichtman}, \citenamefont {Sun},
  \citenamefont {Ebert},\ and\ \citenamefont
  {Saffman}}]{PhysRevLett.123.230501}%
  \BibitemOpen
  \bibfield  {author} {\bibinfo {author} {\bibfnamefont {T.~M.}\ \bibnamefont
  {Graham}}, \bibinfo {author} {\bibfnamefont {M.}~\bibnamefont {Kwon}},
  \bibinfo {author} {\bibfnamefont {B.}~\bibnamefont {Grinkemeyer}}, \bibinfo
  {author} {\bibfnamefont {Z.}~\bibnamefont {Marra}}, \bibinfo {author}
  {\bibfnamefont {X.}~\bibnamefont {Jiang}}, \bibinfo {author} {\bibfnamefont
  {M.~T.}\ \bibnamefont {Lichtman}}, \bibinfo {author} {\bibfnamefont
  {Y.}~\bibnamefont {Sun}}, \bibinfo {author} {\bibfnamefont {M.}~\bibnamefont
  {Ebert}}, \ and\ \bibinfo {author} {\bibfnamefont {M.}~\bibnamefont
  {Saffman}},\ }\href {\doibase 10.1103/PhysRevLett.123.230501} {\bibfield
  {journal} {\bibinfo  {journal} {Phys. Rev. Lett.}\ }\textbf {\bibinfo
  {volume} {123}},\ \bibinfo {pages} {230501} (\bibinfo {year}
  {2019})}\BibitemShut {NoStop}%
\bibitem [{\citenamefont {Sun}\ \emph {et~al.}(2020)\citenamefont {Sun},
  \citenamefont {Xu}, \citenamefont {Chen},\ and\ \citenamefont
  {Liu}}]{PhysRevApplied.13.024059}%
  \BibitemOpen
  \bibfield  {author} {\bibinfo {author} {\bibfnamefont {Y.}~\bibnamefont
  {Sun}}, \bibinfo {author} {\bibfnamefont {P.}~\bibnamefont {Xu}}, \bibinfo
  {author} {\bibfnamefont {P.-X.}\ \bibnamefont {Chen}}, \ and\ \bibinfo
  {author} {\bibfnamefont {L.}~\bibnamefont {Liu}},\ }\href {\doibase
  10.1103/PhysRevApplied.13.024059} {\bibfield  {journal} {\bibinfo  {journal}
  {Phys. Rev. Applied}\ }\textbf {\bibinfo {volume} {13}},\ \bibinfo {pages}
  {024059} (\bibinfo {year} {2020})}\BibitemShut {NoStop}%
\bibitem [{\citenamefont {Jau}\ \emph {et~al.}(2016)\citenamefont {Jau},
  \citenamefont {Hankin}, \citenamefont {Keating}, \citenamefont {Deutsch},\
  and\ \citenamefont {Biedermann}}]{jau2016entangling}%
  \BibitemOpen
  \bibfield  {author} {\bibinfo {author} {\bibfnamefont {Y.-Y.}\ \bibnamefont
  {Jau}}, \bibinfo {author} {\bibfnamefont {A.}~\bibnamefont {Hankin}},
  \bibinfo {author} {\bibfnamefont {T.}~\bibnamefont {Keating}}, \bibinfo
  {author} {\bibfnamefont {I.}~\bibnamefont {Deutsch}}, \ and\ \bibinfo
  {author} {\bibfnamefont {G.}~\bibnamefont {Biedermann}},\ }\href {\doibase
  10.1038/nphys3487} {\bibfield  {journal} {\bibinfo  {journal} {Nat. Phys.}\
  }\textbf {\bibinfo {volume} {12}},\ \bibinfo {pages} {71} (\bibinfo {year}
  {2016})}\BibitemShut {NoStop}%
\bibitem [{\citenamefont {Walker}\ and\ \citenamefont
  {Saffman}(2008)}]{PhysRevA.77.032723}%
  \BibitemOpen
  \bibfield  {author} {\bibinfo {author} {\bibfnamefont {T.~G.}\ \bibnamefont
  {Walker}}\ and\ \bibinfo {author} {\bibfnamefont {M.}~\bibnamefont
  {Saffman}},\ }\href {\doibase 10.1103/PhysRevA.77.032723} {\bibfield
  {journal} {\bibinfo  {journal} {Phys. Rev. A}\ }\textbf {\bibinfo {volume}
  {77}},\ \bibinfo {pages} {032723} (\bibinfo {year} {2008})}\BibitemShut
  {NoStop}%
\bibitem [{\citenamefont {Bernien}\ \emph {et~al.}(2017)\citenamefont
  {Bernien}, \citenamefont {Schwartz}, \citenamefont {Keesling}, \citenamefont
  {Levine}, \citenamefont {Omran}, \citenamefont {Pichler}, \citenamefont
  {Choi}, \citenamefont {Zibrov}, \citenamefont {Endres}, \citenamefont
  {Greiner}, \citenamefont {Vuletić},\ and\ \citenamefont
  {Lukin}}]{Bernien2016}%
  \BibitemOpen
  \bibfield  {author} {\bibinfo {author} {\bibfnamefont {H.}~\bibnamefont
  {Bernien}}, \bibinfo {author} {\bibfnamefont {S.}~\bibnamefont {Schwartz}},
  \bibinfo {author} {\bibfnamefont {A.}~\bibnamefont {Keesling}}, \bibinfo
  {author} {\bibfnamefont {H.}~\bibnamefont {Levine}}, \bibinfo {author}
  {\bibfnamefont {A.}~\bibnamefont {Omran}}, \bibinfo {author} {\bibfnamefont
  {H.}~\bibnamefont {Pichler}}, \bibinfo {author} {\bibfnamefont
  {S.}~\bibnamefont {Choi}}, \bibinfo {author} {\bibfnamefont {A.~S.}\
  \bibnamefont {Zibrov}}, \bibinfo {author} {\bibfnamefont {M.}~\bibnamefont
  {Endres}}, \bibinfo {author} {\bibfnamefont {M.}~\bibnamefont {Greiner}},
  \bibinfo {author} {\bibfnamefont {V.}~\bibnamefont {Vuletić}}, \ and\
  \bibinfo {author} {\bibfnamefont {M.~D.}\ \bibnamefont {Lukin}},\ }\href
  {\doibase 10.1038/nature24622} {\bibfield  {journal} {\bibinfo  {journal}
  {Nature}\ }\textbf {\bibinfo {volume} {551}},\ \bibinfo {pages} {579}
  (\bibinfo {year} {2017})}\BibitemShut {NoStop}%
\bibitem [{\citenamefont {Zhang}\ \emph {et~al.}(2010)\citenamefont {Zhang},
  \citenamefont {Isenhower}, \citenamefont {Gill}, \citenamefont {Walker},\
  and\ \citenamefont {Saffman}}]{PhysRevA.82.030306}%
  \BibitemOpen
  \bibfield  {author} {\bibinfo {author} {\bibfnamefont {X.~L.}\ \bibnamefont
  {Zhang}}, \bibinfo {author} {\bibfnamefont {L.}~\bibnamefont {Isenhower}},
  \bibinfo {author} {\bibfnamefont {A.~T.}\ \bibnamefont {Gill}}, \bibinfo
  {author} {\bibfnamefont {T.~G.}\ \bibnamefont {Walker}}, \ and\ \bibinfo
  {author} {\bibfnamefont {M.}~\bibnamefont {Saffman}},\ }\href {\doibase
  10.1103/PhysRevA.82.030306} {\bibfield  {journal} {\bibinfo  {journal} {Phys.
  Rev. A}\ }\textbf {\bibinfo {volume} {82}},\ \bibinfo {pages} {030306(R)}
  (\bibinfo {year} {2010})}\BibitemShut {NoStop}%
\bibitem [{\citenamefont {Isenhower}\ \emph {et~al.}(2010)\citenamefont
  {Isenhower}, \citenamefont {Urban}, \citenamefont {Zhang}, \citenamefont
  {Gill}, \citenamefont {Henage}, \citenamefont {Johnson}, \citenamefont
  {Walker},\ and\ \citenamefont {Saffman}}]{PhysRevLett.104.010503}%
  \BibitemOpen
  \bibfield  {author} {\bibinfo {author} {\bibfnamefont {L.}~\bibnamefont
  {Isenhower}}, \bibinfo {author} {\bibfnamefont {E.}~\bibnamefont {Urban}},
  \bibinfo {author} {\bibfnamefont {X.~L.}\ \bibnamefont {Zhang}}, \bibinfo
  {author} {\bibfnamefont {A.~T.}\ \bibnamefont {Gill}}, \bibinfo {author}
  {\bibfnamefont {T.}~\bibnamefont {Henage}}, \bibinfo {author} {\bibfnamefont
  {T.~A.}\ \bibnamefont {Johnson}}, \bibinfo {author} {\bibfnamefont {T.~G.}\
  \bibnamefont {Walker}}, \ and\ \bibinfo {author} {\bibfnamefont
  {M.}~\bibnamefont {Saffman}},\ }\href {\doibase
  10.1103/PhysRevLett.104.010503} {\bibfield  {journal} {\bibinfo  {journal}
  {Phys. Rev. Lett.}\ }\textbf {\bibinfo {volume} {104}},\ \bibinfo {pages}
  {010503} (\bibinfo {year} {2010})}\BibitemShut {NoStop}%
\bibitem [{\citenamefont {Wilk}\ \emph {et~al.}(2010)\citenamefont {Wilk},
  \citenamefont {Ga\"etan}, \citenamefont {Evellin}, \citenamefont {Wolters},
  \citenamefont {Miroshnychenko}, \citenamefont {Grangier},\ and\ \citenamefont
  {Browaeys}}]{PhysRevLett.104.010502}%
  \BibitemOpen
  \bibfield  {author} {\bibinfo {author} {\bibfnamefont {T.}~\bibnamefont
  {Wilk}}, \bibinfo {author} {\bibfnamefont {A.}~\bibnamefont {Ga\"etan}},
  \bibinfo {author} {\bibfnamefont {C.}~\bibnamefont {Evellin}}, \bibinfo
  {author} {\bibfnamefont {J.}~\bibnamefont {Wolters}}, \bibinfo {author}
  {\bibfnamefont {Y.}~\bibnamefont {Miroshnychenko}}, \bibinfo {author}
  {\bibfnamefont {P.}~\bibnamefont {Grangier}}, \ and\ \bibinfo {author}
  {\bibfnamefont {A.}~\bibnamefont {Browaeys}},\ }\href {\doibase
  10.1103/PhysRevLett.104.010502} {\bibfield  {journal} {\bibinfo  {journal}
  {Phys. Rev. Lett.}\ }\textbf {\bibinfo {volume} {104}},\ \bibinfo {pages}
  {010502} (\bibinfo {year} {2010})}\BibitemShut {NoStop}%
\bibitem [{\citenamefont {Beterov}\ \emph {et~al.}(2009)\citenamefont
  {Beterov}, \citenamefont {Ryabtsev}, \citenamefont {Tretyakov},\ and\
  \citenamefont {Entin}}]{PhysRevA.79.052504}%
  \BibitemOpen
  \bibfield  {author} {\bibinfo {author} {\bibfnamefont {I.~I.}\ \bibnamefont
  {Beterov}}, \bibinfo {author} {\bibfnamefont {I.~I.}\ \bibnamefont
  {Ryabtsev}}, \bibinfo {author} {\bibfnamefont {D.~B.}\ \bibnamefont
  {Tretyakov}}, \ and\ \bibinfo {author} {\bibfnamefont {V.~M.}\ \bibnamefont
  {Entin}},\ }\href {\doibase 10.1103/PhysRevA.79.052504} {\bibfield  {journal}
  {\bibinfo  {journal} {Phys. Rev. A}\ }\textbf {\bibinfo {volume} {79}},\
  \bibinfo {pages} {052504} (\bibinfo {year} {2009})}\BibitemShut {NoStop}%
\bibitem [{\citenamefont {Fowler}\ \emph {et~al.}(2009)\citenamefont {Fowler},
  \citenamefont {Stephens},\ and\ \citenamefont
  {Groszkowski}}]{PhysRevA.80.052312}%
  \BibitemOpen
  \bibfield  {author} {\bibinfo {author} {\bibfnamefont {A.~G.}\ \bibnamefont
  {Fowler}}, \bibinfo {author} {\bibfnamefont {A.~M.}\ \bibnamefont
  {Stephens}}, \ and\ \bibinfo {author} {\bibfnamefont {P.}~\bibnamefont
  {Groszkowski}},\ }\href {\doibase 10.1103/PhysRevA.80.052312} {\bibfield
  {journal} {\bibinfo  {journal} {Phys. Rev. A}\ }\textbf {\bibinfo {volume}
  {80}},\ \bibinfo {pages} {052312} (\bibinfo {year} {2009})}\BibitemShut
  {NoStop}%
\bibitem [{\citenamefont {Keating}\ \emph {et~al.}(2015)\citenamefont
  {Keating}, \citenamefont {Cook}, \citenamefont {Hankin}, \citenamefont {Jau},
  \citenamefont {Biedermann},\ and\ \citenamefont
  {Deutsch}}]{PhysRevA.91.012337}%
  \BibitemOpen
  \bibfield  {author} {\bibinfo {author} {\bibfnamefont {T.}~\bibnamefont
  {Keating}}, \bibinfo {author} {\bibfnamefont {R.~L.}\ \bibnamefont {Cook}},
  \bibinfo {author} {\bibfnamefont {A.~M.}\ \bibnamefont {Hankin}}, \bibinfo
  {author} {\bibfnamefont {Y.-Y.}\ \bibnamefont {Jau}}, \bibinfo {author}
  {\bibfnamefont {G.~W.}\ \bibnamefont {Biedermann}}, \ and\ \bibinfo {author}
  {\bibfnamefont {I.~H.}\ \bibnamefont {Deutsch}},\ }\href {\doibase
  10.1103/PhysRevA.91.012337} {\bibfield  {journal} {\bibinfo  {journal} {Phys.
  Rev. A}\ }\textbf {\bibinfo {volume} {91}},\ \bibinfo {pages} {012337}
  (\bibinfo {year} {2015})}\BibitemShut {NoStop}%
\bibitem [{\citenamefont {Saffman}\ \emph {et~al.}(2011)\citenamefont
  {Saffman}, \citenamefont {Zhang}, \citenamefont {Gill}, \citenamefont
  {Isenhower},\ and\ \citenamefont {Walker}}]{Saffman_2011}%
  \BibitemOpen
  \bibfield  {author} {\bibinfo {author} {\bibfnamefont {M.}~\bibnamefont
  {Saffman}}, \bibinfo {author} {\bibfnamefont {X.~L.}\ \bibnamefont {Zhang}},
  \bibinfo {author} {\bibfnamefont {A.~T.}\ \bibnamefont {Gill}}, \bibinfo
  {author} {\bibfnamefont {L.}~\bibnamefont {Isenhower}}, \ and\ \bibinfo
  {author} {\bibfnamefont {T.~G.}\ \bibnamefont {Walker}},\ }\href {\doibase
  10.1088/1742-6596/264/1/012023} {\bibfield  {journal} {\bibinfo  {journal}
  {J. Phys. Conf. Ser.}\ }\textbf {\bibinfo {volume} {264}},\ \bibinfo {pages}
  {012023} (\bibinfo {year} {2011})}\BibitemShut {NoStop}%
\bibitem [{\citenamefont {Saffman}(2016)}]{Saffman_2016}%
  \BibitemOpen
  \bibfield  {author} {\bibinfo {author} {\bibfnamefont {M.}~\bibnamefont
  {Saffman}},\ }\href {\doibase 10.1088/0953-4075/49/20/202001} {\bibfield
  {journal} {\bibinfo  {journal} {J. Phys. B: At. Mol. Opt. Phys.}\ }\textbf
  {\bibinfo {volume} {49}},\ \bibinfo {pages} {202001} (\bibinfo {year}
  {2016})}\BibitemShut {NoStop}%
\bibitem [{\citenamefont {Ryabtsev}\ \emph {et~al.}(2011)\citenamefont
  {Ryabtsev}, \citenamefont {Beterov}, \citenamefont {Tretyakov}, \citenamefont
  {Entin},\ and\ \citenamefont {Yakshina}}]{PhysRevA.84.053409}%
  \BibitemOpen
  \bibfield  {author} {\bibinfo {author} {\bibfnamefont {I.~I.}\ \bibnamefont
  {Ryabtsev}}, \bibinfo {author} {\bibfnamefont {I.~I.}\ \bibnamefont
  {Beterov}}, \bibinfo {author} {\bibfnamefont {D.~B.}\ \bibnamefont
  {Tretyakov}}, \bibinfo {author} {\bibfnamefont {V.~M.}\ \bibnamefont
  {Entin}}, \ and\ \bibinfo {author} {\bibfnamefont {E.~A.}\ \bibnamefont
  {Yakshina}},\ }\href {\doibase 10.1103/PhysRevA.84.053409} {\bibfield
  {journal} {\bibinfo  {journal} {Phys. Rev. A}\ }\textbf {\bibinfo {volume}
  {84}},\ \bibinfo {pages} {053409} (\bibinfo {year} {2011})}\BibitemShut
  {NoStop}%
\bibitem [{\citenamefont {Shi}(2020)}]{PhysRevApplied.13.024008}%
  \BibitemOpen
  \bibfield  {author} {\bibinfo {author} {\bibfnamefont {X.-F.}\ \bibnamefont
  {Shi}},\ }\href {\doibase 10.1103/PhysRevApplied.13.024008} {\bibfield
  {journal} {\bibinfo  {journal} {Phys. Rev. Applied}\ }\textbf {\bibinfo
  {volume} {13}},\ \bibinfo {pages} {024008} (\bibinfo {year}
  {2020})}\BibitemShut {NoStop}%
\bibitem [{\citenamefont {Sun}\ \emph {et~al.}(2019)\citenamefont {Sun},
  \citenamefont {Xu},\ and\ \citenamefont {Liu}}]{sun2019doppler}%
  \BibitemOpen
  \bibfield  {author} {\bibinfo {author} {\bibfnamefont {Y.}~\bibnamefont
  {Sun}}, \bibinfo {author} {\bibfnamefont {P.}~\bibnamefont {Xu}}, \ and\
  \bibinfo {author} {\bibfnamefont {L.}~\bibnamefont {Liu}},\ }\href@noop {}
  {\enquote {\bibinfo {title} {Doppler-insensitive two-qubit controlled-phase
  gate protocol with dual-pulse off-resonant modulated driving for neutral
  atoms},}\ } (\bibinfo {year} {2019}),\ \Eprint
  {http://arxiv.org/abs/1912.04504} {arXiv:1912.04504} \BibitemShut {NoStop}%
\bibitem [{\citenamefont {Shi}(2018{\natexlab{b}})}]{PhysRevApplied.10.034006}%
  \BibitemOpen
  \bibfield  {author} {\bibinfo {author} {\bibfnamefont {X.-F.}\ \bibnamefont
  {Shi}},\ }\href {\doibase 10.1103/PhysRevApplied.10.034006} {\bibfield
  {journal} {\bibinfo  {journal} {Phys. Rev. Applied}\ }\textbf {\bibinfo
  {volume} {10}},\ \bibinfo {pages} {034006} (\bibinfo {year}
  {2018}{\natexlab{b}})}\BibitemShut {NoStop}%
\bibitem [{\citenamefont {de~L\'es\'eleuc}\ \emph {et~al.}(2018)\citenamefont
  {de~L\'es\'eleuc}, \citenamefont {Barredo}, \citenamefont {Lienhard},
  \citenamefont {Browaeys},\ and\ \citenamefont {Lahaye}}]{PhysRevA.97.053803}%
  \BibitemOpen
  \bibfield  {author} {\bibinfo {author} {\bibfnamefont {S.}~\bibnamefont
  {de~L\'es\'eleuc}}, \bibinfo {author} {\bibfnamefont {D.}~\bibnamefont
  {Barredo}}, \bibinfo {author} {\bibfnamefont {V.}~\bibnamefont {Lienhard}},
  \bibinfo {author} {\bibfnamefont {A.}~\bibnamefont {Browaeys}}, \ and\
  \bibinfo {author} {\bibfnamefont {T.}~\bibnamefont {Lahaye}},\ }\href
  {\doibase 10.1103/PhysRevA.97.053803} {\bibfield  {journal} {\bibinfo
  {journal} {Phys. Rev. A}\ }\textbf {\bibinfo {volume} {97}},\ \bibinfo
  {pages} {053803} (\bibinfo {year} {2018})}\BibitemShut {NoStop}%
\bibitem [{\citenamefont {Stiesdal}\ \emph {et~al.}(2018)\citenamefont
  {Stiesdal}, \citenamefont {Kumlin}, \citenamefont {Kleinbeck}, \citenamefont
  {Lunt}, \citenamefont {Braun}, \citenamefont {Paris-Mandoki}, \citenamefont
  {Tresp}, \citenamefont {B\"uchler},\ and\ \citenamefont
  {Hofferberth}}]{PhysRevLett.121.103601}%
  \BibitemOpen
  \bibfield  {author} {\bibinfo {author} {\bibfnamefont {N.}~\bibnamefont
  {Stiesdal}}, \bibinfo {author} {\bibfnamefont {J.}~\bibnamefont {Kumlin}},
  \bibinfo {author} {\bibfnamefont {K.}~\bibnamefont {Kleinbeck}}, \bibinfo
  {author} {\bibfnamefont {P.}~\bibnamefont {Lunt}}, \bibinfo {author}
  {\bibfnamefont {C.}~\bibnamefont {Braun}}, \bibinfo {author} {\bibfnamefont
  {A.}~\bibnamefont {Paris-Mandoki}}, \bibinfo {author} {\bibfnamefont
  {C.}~\bibnamefont {Tresp}}, \bibinfo {author} {\bibfnamefont {H.~P.}\
  \bibnamefont {B\"uchler}}, \ and\ \bibinfo {author} {\bibfnamefont
  {S.}~\bibnamefont {Hofferberth}},\ }\href {\doibase
  10.1103/PhysRevLett.121.103601} {\bibfield  {journal} {\bibinfo  {journal}
  {Phys. Rev. Lett.}\ }\textbf {\bibinfo {volume} {121}},\ \bibinfo {pages}
  {103601} (\bibinfo {year} {2018})}\BibitemShut {NoStop}%
\bibitem [{\citenamefont {Zhao}\ \emph {et~al.}(2017)\citenamefont {Zhao},
  \citenamefont {Liu}, \citenamefont {Ji}, \citenamefont {Tang},\ and\
  \citenamefont {Shao}}]{Zhao2017}%
  \BibitemOpen
  \bibfield  {author} {\bibinfo {author} {\bibfnamefont {Y.~J.}\ \bibnamefont
  {Zhao}}, \bibinfo {author} {\bibfnamefont {B.}~\bibnamefont {Liu}}, \bibinfo
  {author} {\bibfnamefont {Y.~Q.}\ \bibnamefont {Ji}}, \bibinfo {author}
  {\bibfnamefont {S.~Q.}\ \bibnamefont {Tang}}, \ and\ \bibinfo {author}
  {\bibfnamefont {X.~Q.}\ \bibnamefont {Shao}},\ }\href {\doibase
  10.1038/s41598-017-16533-9} {\bibfield  {journal} {\bibinfo  {journal} {Sci.
  Rep.}\ }\textbf {\bibinfo {volume} {7}},\ \bibinfo {pages} {16489} (\bibinfo
  {year} {2017})}\BibitemShut {NoStop}%
\bibitem [{\citenamefont {Han}\ \emph {et~al.}(2020)\citenamefont {Han},
  \citenamefont {Wu}, \citenamefont {Wang}, \citenamefont {Jiang},
  \citenamefont {Xia},\ and\ \citenamefont {Song}}]{Han:20}%
  \BibitemOpen
  \bibfield  {author} {\bibinfo {author} {\bibfnamefont {J.-X.}\ \bibnamefont
  {Han}}, \bibinfo {author} {\bibfnamefont {J.-L.}\ \bibnamefont {Wu}},
  \bibinfo {author} {\bibfnamefont {Y.}~\bibnamefont {Wang}}, \bibinfo {author}
  {\bibfnamefont {Y.-Y.}\ \bibnamefont {Jiang}}, \bibinfo {author}
  {\bibfnamefont {Y.}~\bibnamefont {Xia}}, \ and\ \bibinfo {author}
  {\bibfnamefont {J.}~\bibnamefont {Song}},\ }\href {\doibase
  10.1364/OE.384352} {\bibfield  {journal} {\bibinfo  {journal} {Opt. Express}\
  }\textbf {\bibinfo {volume} {28}},\ \bibinfo {pages} {1954} (\bibinfo {year}
  {2020})}\BibitemShut {NoStop}%
\bibitem [{\citenamefont {Veps{\"a}l{\"a}inen}\ \emph
  {et~al.}(2019)\citenamefont {Veps{\"a}l{\"a}inen}, \citenamefont {Danilin},\
  and\ \citenamefont {Paraoanu}}]{Vepsaaineneaau5999}%
  \BibitemOpen
  \bibfield  {author} {\bibinfo {author} {\bibfnamefont {A.}~\bibnamefont
  {Veps{\"a}l{\"a}inen}}, \bibinfo {author} {\bibfnamefont {S.}~\bibnamefont
  {Danilin}}, \ and\ \bibinfo {author} {\bibfnamefont {G.~S.}\ \bibnamefont
  {Paraoanu}},\ }\href {\doibase 10.1126/sciadv.aau5999} {\bibfield  {journal}
  {\bibinfo  {journal} {Sci. Adv.}\ }\textbf {\bibinfo {volume} {5}},\ \bibinfo
  {pages} {eaau5999} (\bibinfo {year} {2019})}\BibitemShut {NoStop}%
\end{thebibliography}%
\end{document}